\documentclass[10pt]{iopart} 
\usepackage{iopams}
\usepackage{bm,amssymb,setstack}
\usepackage{graphicx,caption,subcaption}

\usepackage[colorlinks=true,citecolor=blue]{hyperref}

\newcommand{\eqref}[1]{{(\ref{#1})}}    %simpler referencing

\newcommand{\bx}{{\bm x}}
\newcommand{\bX}{{\bm X}}

\newcommand{\bg}{{\bm \gamma}}

\newcommand{\bH}{{\bm H}}

\begin{document}
\title{Optics in a nonlinear gravitational plane wave}
\author{Abraham I. Harte}
\address{Max-Planck-Institut f\"{u}r Gravitationsphysik, Albert-Einstein-Institut
	\\
	 Am M\"{u}hlenberg 1, 14476 Golm, Germany}

\begin{abstract}
Gravitational waves can act like gravitational lenses, affecting the observed positions, brightnesses, and redshifts of distant objects. Exact expressions for such effects are derived here in general relativity, allowing for arbitrarily-moving sources and observers in the presence of plane-symmetric gravitational waves. At least for freely falling sources and observers, it is shown that the commonly-used predictions of linear perturbation theory can be generically overshadowed by nonlinear effects; even for very weak gravitational waves, higher-order perturbative corrections involve secularly-growing terms which cannot necessarily be neglected when considering observations of sufficiently distant sources. Even on more moderate scales where linear effects remain at least marginally dominant, nonlinear corrections are qualitatively different from their linear counterparts. There is a sense in which they can, for example, mimic the existence of a third type of gravitational wave polarization. 
\end{abstract}

\section{Introduction}

Some of the most important potential signatures of gravitational waves are associated with their effects on the propagation of light. Collections of null rays can be deflected, sheared, delayed, or otherwise altered as they travel through a gravitational wave. Indeed, most contemporary attempts to observe gravitational waves rely on measurements of the relative time delays which accumulate as light travels between material bodies. This is particularly clear for interferometric detectors \cite{InterferometerReview}, where one or more beams of light are circulated between collections of mirrors and then recombined to reveal their relative phases. Efforts to detect gravitational waves using pulsar timing arrays \cite{PTAreview} exploit similar principles, but instead make use of time intervals observed on the earth between radio bursts emitted by distant pulsars. Besides temporal effects such as these, gravitational waves can also affect observations of an object's sky location, brightness, shape, and so on \cite{HarteLensing, BookFlanagan, Zipoy, DamourBending, Kopeikin, FaraoniRotation, FaraoniRedshift}.

Almost all prior discussions of these phenomena have been perturbative, involving calculations which are valid only through first order in the gravitational wave amplitude (see, however, \cite{HarteLensing, Helfer, Finn2, PerlickReview, Bicak}). This has been justified, at least implicitly, by the minuscule size of even these lowest-order terms: In most cases of astrophysical interest, the gravitational wave strain amplitude $\epsilon$ is much smaller than unity. Enormous technological effort is required to detect such waves at all, and waveform measurements which are accurate to more than a handful of decimal places cannot be expected for quite some time. In this context, it might appear reasonable to dismiss higher-order corrections as uninterestingly-small. One of the goals of this paper is to show that such reasoning can be misleading. Even if a dimensionless observable associated with a gravitational wave of amplitude $\epsilon \ll 1$ is bounded by $\epsilon \times (\mbox{number of order $1$})$ in linear perturbation theory, higher-order corrections are not necessarily bounded by $\epsilon^2 \times (\mbox{another number of order $1$})$. The coefficient in front of the $\epsilon^2$ term can instead grow enormously with the distance between a light source and its observer, implying that nonlinearities may be significant even when considering observations of very weak gravitational waves. Nonlinear effects also tend to have very different observational signatures from their lower-order counterparts, further increasing their potential detectability. 

Although it does not appear to have been previously pointed out in this context, the existence of large higher-order corrections is well-known in many types of perturbative calculations. A simple example is provided by the Mathieu-type equation
\begin{equation}
  \ddot{\xi}(u) + \frac{1}{2} \epsilon \xi(u) \cos u = 0.
  \label{Mathieu}
\end{equation}
If $u$ denotes an appropriate phase coordinate, $\xi^2(u)$ may be shown to describe a particular metric component associated with a linearly-polarized, ``monochromatic'' gravitational plane wave with strain amplitude $\epsilon$. Moreover, the coordinate system where this is true is constructed such that there is a sense in which electromagnetic observations of distant objects have properties which can be read off directly from $\xi^2(u)$. Solutions to \eqref{Mathieu} therefore serve as a convenient proxy for understanding nonlinear effects associated with monochromatic gravitational waves in general relativity. Assuming $\epsilon \ll 1$ while adopting convenient initial conditions,
\begin{equation}
  \xi^2 (u) = 1 + \epsilon \cos u + \frac{1}{8} \epsilon^2 (3 \cos^2 u - u^2) + O(\epsilon^3).
  \label{MathieuExpand}
\end{equation}
The magnitude of the second-order term in this expansion clearly overtakes the first when $|u| \sim \epsilon^{-1/2} \gg 1$, signaling that the linear approximation fails for large $|u|$. This occurs no matter how small $\epsilon$ may be; weaker amplitudes merely delay such problems to larger scales. 

\begin{figure}
    \centering
    \includegraphics[width=.6 \linewidth]{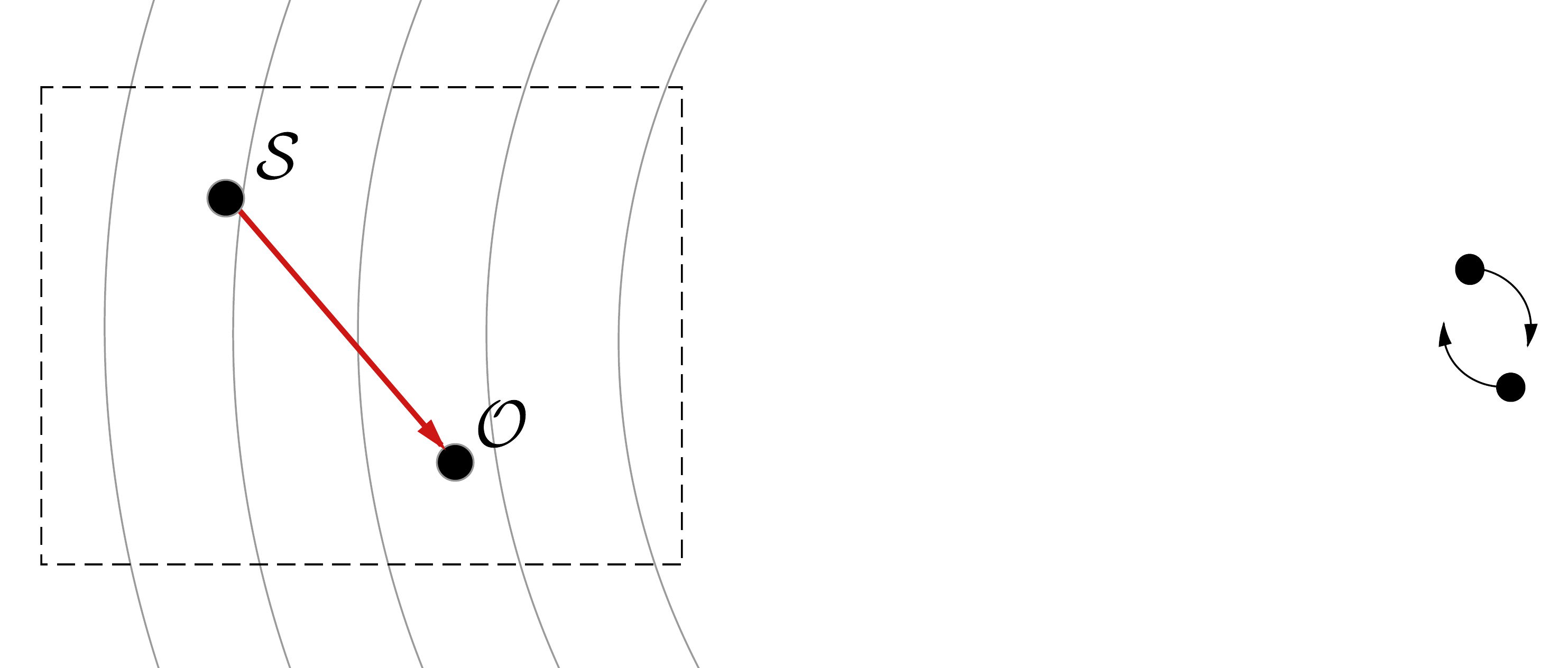}
    \caption{Schematic of a physical system which could correspond to the model  considered in this paper. A ``source'' $\mathcal{S}$ emits electromagnetic radiation which is viewed by an ``observer'' $\mathcal{O}$. In between these objects, spacetime is assumed to be approximately flat except for a nearly-planar gravitational wave. This wave may be generated by a distant binary, although all considerations here are restricted to the boxed region, and are therefore indifferent to the precise nature of the wave generation mechanism.}
    \label{Fig:setup}
\end{figure}

We show that similar effects arise for a variety of gravitational wave observables in general relativity, thus implying that the results of linear perturbation theory cannot necessarily be applied on large scales. As a model, geometric optics is considered in the presence of a plane-symmetric gravitational wave. This may be viewed as an idealization of the system illustrated in Figure \ref{Fig:setup}, where observations are performed sufficiently far from a gravitational wave source that the curvature of the wavefronts may be neglected. Similar models are common (though restricted only to first-order metric perturbations) in, e.g., descriptions for how gravitational waves can affect pulsar timing measurements \cite{PTAreview, Detweiler}.

Despite this, real astrophysical observations cannot rely solely upon plane wave calculations. Deviations from planarity, waves propagating in multiple directions, non-radiative metric perturbations, and other effects could all be significant in observationally-relevant systems. Although calculations which take into account many such effects have been performed through first post-Minkowskian order \cite{Kopeikin, Kopeikin2}, the optical characteristics of \textit{nonlinear} gravitational waves have been essentially unexplored in this context. Moreover, it would likely be difficult to understand the implications of any such calculations even if they did exist; the known first-order expressions are already extremely complicated in their most general forms. Plane waves are, by contrast, sufficiently simple that their physical effects can be thoroughly explored even in the nonlinear regime. At the same time, these waves remain sufficiently complicated to be interesting, and also to capture much of the relevant physics. Results obtained using plane wave descriptions may therefore be useful in the construction of specific hypotheses whose generality can later be tested using more complicated models. The technical details of the plane wave problem can also be used to suggest potential simplifications in more general calculations.

A completely separate motivation for considering plane wave spacetimes follows from a mathematical device known as the Penrose limit \cite{PenroseLimit, BlauPenrose, HarteCaustics}. This provides a sense by which the geometry near any null geodesic in any spacetime is equivalent to the geometry of an appropriate plane wave. It can be interpreted as a statement that the metric in a small region around any sufficiently-relativistic observer in any spacetime is equivalent to that of an ``effective plane wave.'' Although we make no attempt to prove it, the Penrose limit suggests that (at least some types of) observations performed by ultrarelativistic observers can \textit{in general} be reduced to analogous observations in effective plane wave spacetimes.

Section \ref{Sect:ApproxAndExact} reviews gravitational plane waves in general relativity, first from the viewpoint of perturbation theory, and then as exact solutions to Einstein's equation. Relations between these two perspectives and their relative advantages are described in detail. Next, Section \ref{Sect:Optics} considers the physical consequences of plane wave spacetimes by deriving exact time delays, frequency shifts, observed sky positions, area distances, and luminosity distances. With appropriate identifications, some of the resulting expressions are only slightly more complicated than their linearized counterparts. Formal perturbative expansions are nevertheless derived in Section \ref{Sect:Pert} and then applied to specific examples in Section \ref{Sect:Example}.

\subsection*{Notation and conventions}

The metric signature here is $+2$, $c=G=1$, the Riemann tensor satisfies $R_{abc}{}^{d} \omega_d = 2 \nabla_{[a} \nabla_{b]} \omega_c$ for any $\omega_c$, Latin letters $a,b,\ldots$ denote abstract indices, Greek letters $\mu,\nu,\ldots$ denote four-dimensional coordinate indices, and $i,j, \ldots$ are used as two-dimensional coordinate indices associated with directions transverse to the background gravitational wave. When convenient, transverse coordinate components are indicated using boldface symbols without indices [e.g., $\gamma_{ij} = (\bg)_{ij}$, $\gamma_{ij} w^i v^j = \bm{w}^\intercal \bg \bm{v}$, and $\tr \bg = \delta_{ij} \gamma_{ij}$]. The one exception where the boldface symbol doesn't correspond to its component counterpart is the $2 \times 2$ identity matrix $(\bm{I})_{ij} = \delta_{ij}$. Lastly, overdots are used to denote derivatives with respect to a phase coordinate $u$, so $\dot{\gamma}_{ij} = \rmd \gamma_{ij}/\rmd u$.

\section{Gravitational plane waves}
\label{Sect:ApproxAndExact}

Before describing how light propagates in a gravitational plane wave, it must first be explained precisely what a plane wave is. Although the concept is clear for a scalar field in flat spacetime, subtleties arise when considering curved geometries or fields with nontrivial tensorial structure. This difficulty is reflected in part by the two distinct perspectives on gravitational plane waves---one perturbative and one exact---which are common in the literature. While both of these perspectives are individually well-known, the relations between them are not. This section clarifies the situation, and also remarks on special types of plane waves.

\subsection{Approximate plane waves}

Almost every textbook on general relativity discusses gravitational waves as linear perturbations on a flat background spacetime \cite{MTW,Wald}. To review, suppose that there exist coordinates $(t,x^1,x^2,z)$ in which the metric components can be approximated by $g_{\mu\nu} = \eta_{\mu\nu} + \epsilon h_{\mu\nu} + O(\epsilon^2)$, where $\epsilon \ll 1$ is a dimensionless expansion parameter and $\eta_{\mu\nu}$ is the Minkowski metric. It is also typical to adopt the transverse-traceless, or ``TT'' gauge, in which case $h_{\mu t}  = \eta^{\mu\nu} h_{\mu\nu}  = \eta^{\mu\nu} \partial_\mu h_{\nu\lambda} = 0$. These constraints on the metric perturbation can always be imposed in connected regions of spacetime where the linearized vacuum Einstein equation holds \cite{FlanaganHughes}. Imposing  TT gauge in such a region, the first-order vacuum Einstein equation reduces there to the ordinary flat-spacetime wave equation for each coordinate component of the perturbation: $(-\partial_t^2 + \nabla^2 ) h_{\mu\nu} = 0$.

The solutions of interest here represent plane-symmetric gravitational waves propagating in vacuum. If the spatial projection of a plane wave's direction of propagation is identified with $\partial/\partial z$, the only components of the metric perturbation which might not vanish are $h_{ij}$, where $i=1,2$. This $2 \times 2$ symmetric matrix must be trace-free, and can depend only on the ``phase coordinate'' $u = (t-z)/\sqrt{2}$. The first-order line element for an arbitrary plane wave propagating in the $z$-direction is therefore
\begin{equation}
    \rmd s^2 = - \rmd t^2 + [\delta_{ij} + \epsilon h_{ij} (u)] \rmd x^i \rmd x^j + \rmd z^2 + O (\epsilon^2) ,
    \label{TTgauge}
\end{equation}
where 
\begin{equation}
    \bm{h}(u) = 
        \left(
        \begin{array}{cc}
          h_{+}(u) & h_{\times}(u) \\
          h_{\times}(u) & -h_{+}(u)
        \end{array}
        \right),
    \label{hPlane}
\end{equation}
and $h_+(u)$, $h_\times(u)$ represent the waveforms associated with the ``$+$'' and ``$\times$'' polarization states. A plane wave is said to be linearly polarized if $h_\times$ (or $h_+$) can be made to vanish via some constant rotation of the $x^1$, $x^2$ coordinates. 

\subsection{Exact plane waves}
\label{Sect:Brinkmann}

The exact theory of gravitational plane waves is often presented very differently from its linearized counterpart. One way to motivate an exact plane wave solution in general relativity is to first search for those geometries which share---independently of Einstein's equation---the symmetries associated with  more familiar plane waves in flat spacetime. Consider, for example, a scalar field with the form $f = f(t-z)$ in special relativity. This is clearly symmetric with respect to the two spacelike translations $\partial/\partial x^i$ and the single null translation $\partial/\partial t + \partial/ \partial z$. Less obviously, scalar plane waves are also preserved by
\begin{equation}
  \left( x^i \frac{\partial}{\partial z} - z \frac{\partial}{\partial x^i} \right) + \left( x^i \frac{\partial}{\partial t} + t \frac{\partial}{\partial x^i} \right),
  \label{SpecialKilling}
\end{equation}
two Killing vector fields which generate rotations in the $x^i$-$z$ planes combined with boosts along the $x^i$ directions. It follows that scalar plane waves in flat spacetime are preserved by at least five Killing vector fields associated with the geometry through which they propagate. The same symmetries also preserve electromagnetic plane waves in flat spacetime, thus motivating a gravitational plane wave in four dimensions as a curved spacetime which admits at least five linearly-independent Killing vector fields \cite{Bondi}. The resulting metrics are most commonly-stated in terms of the so-called Brinkmann coordinates $(U,V,X^1,X^2)$, in which case
\begin{equation}
    \rmd s^2= - 2 \rmd U \rmd V + H_{ij}(U) X^i X^j \rmd U^2 + (\rmd X^1)^2 +(\rmd X^2)^2.
    \label{Brinkmann}
\end{equation}
Using this as an ansatz, the \textit{exact} vacuum Einstein equation reduces to
\begin{equation}
  \tr \bH = 0.
  \label{TrH}
\end{equation}
Any $2 \times 2$ symmetric trace-free matrix $H_{ij}(U)$ therefore describes an exact plane wave in vacuum general relativity. We call this matrix the Brinkmann waveform. It has a direct geometric significance in the sense that the only independent, non-vanishing components of the Riemann tensor are
\begin{equation}
R_{UiUj} = - H_{ij},
    \label{Riemann}
\end{equation}
where $i,j$ refer to coordinate components associated with the two $X^i$ coordinates. Furthermore, the two independent components of $H_{ij}$ correspond to the two possible polarization states of a gravitational wave in vacuum general relativity. Up to coordinate ambiguities in the construction of $R_{UiUj}$, it follows from \eqref{Riemann} that the Brinkmann waveform may be obtained using only local measurements. Such ambiguities are minor, and can be taken into account using only three constants $c_b$, $c_r$, and $c_t$ \cite{EhlersKundt}. Explicitly, two waveforms are physically identical if and only if they can be related via the complex replacement rule
\begin{equation}
    H_{11}(U) + i H_{12}(U) \rightarrow c_b^2 [ H_{11}(c_b U - c_t) + i H_{12}(c_b U - c_t)] e^{i c_r},
    \label{Htrans}
\end{equation}
where $c_b \neq 0$ describes a constant boost along the direction of propagation, $c_t$ a constant translation of the phase coordinate $U$, and $c_r$ a constant rotation. This is sufficiently simple that it is typically evident by inspection whether or not two Brinkmann waveforms describe the same physical system.

Physically, a plane wave described by the metric \eqref{Brinkmann} propagates in the null direction $\ell^a = \partial/\partial V$. Up to an overall constant, $\ell^a$ is the unique nonvanishing vector field which is both null and covariantly constant in any curved region of a plane wave spacetime. Noting that $\ell_a = - \nabla_a U$, these constraints define the $U$ coordinate up to an overall affine transformation. It may be interpreted as the phase of the gravitational wave. Interpretations of the remaining Brinkmann coordinates follow by noting that they form a kind of Fermi normal coordinate system whose ``origin'' is the null geodesic $V = X^i = 0$ \cite{BlauPenrose}. 

Plane waves are extremely simple when described in terms of Brinkmann waveforms. It is clear from \eqref{TrH} that if $H_{ij}(U)$ and $H'_{ij}(U)$ are any two vacuum waveforms, $H_{ij}(U) + H'_{ij}(U)$ is also a vacuum waveform. This provides a sense by which plane waves in general relativity satisfy exact linear superposition; there is no nonlinearity to speak of. Superpositions of this type are special cases of the more general result that the vacuum Einstein equation is linear for all metrics within the Kerr-Schild class. More precisely, suppose that there exists some null $\hat{\ell}_a$ and some scalar $\mathcal{H}$ such that $g_{ab} = \hat{g}_{ab} + \mathcal{H} \hat{\ell}_a \hat{\ell}_b$ is an exact solution to the vacuum Einstein equation for some vacuum ``background'' $\hat{g}_{ab}$. If $g_{ab}' = \hat{g}_{ab} + \mathcal{H}' \hat{\ell}_a \hat{\ell}_b$ is a second exact solution, the metric $g''_{ab} = \hat{g}_{ab} + (\mathcal{H} + \mathcal{H}') \hat{\ell}_a \hat{\ell}_b$ must be an exact solution as well \cite{Xanthopoulos}. For the plane wave case of interest here, the Kerr-Schild decomposition is recovered by letting $\hat{\ell}_a = \ell_a = - \nabla_a U$ and $\mathcal{H} = H_{ij} X^i X^j$, in which case $\hat{g}_{ab}$ is flat. Brinkmann waveforms are very special, and the linearity of Einstein's equation which they make manifest is not at all apparent if plane waves are parametrized using different variables. Nevertheless, the optical observables discussed below are more conveniently described in terms of different waveforms which i) generalize the perturbative $h_{ij}$ appearing in \eqref{TTgauge}, and ii) do not satisfy linear superposition.

\subsection{Relating the exact and approximate descriptions}
\label{Sect:BrinkToRosen}

Although the TT-gauge metric \eqref{TTgauge} describes, at least approximately, the same physical system as the Brinkmann metric \eqref{Brinkmann}, the former expression is not a trivial linearization of the latter. Understanding how these two descriptions relate to one another requires an appropriate coordinate transformation. The TT gauge makes sense only at first perturbative order, so a non-perturbative generalization of this gauge must be sought. We choose to employ those transformations which preserve, in an appropriate sense, the TT-gauge property that objects at ``fixed spatial coordinates'' move on timelike geodesics. Additionally, we require that the planar symmetry of the metric be manifest in the sense that there exist two coordinates $x^i$, interpreted as parametrizing the 2-surfaces transverse to the gravitational wave, such that the vector fields $\partial/\partial x^i$ are both spacelike and Killing. These constraints are chosen not merely to facilitate a simple translation between the perturbative and non-perturbative viewpoints. Much more importantly, it is shown in Section \ref{Sect:Optics} below that the ``TT-like'' coordinates which they define are particularly well-adapted to describing non-perturbative optical observables in the presence of an arbitrarily-strong gravitational wave. 

The appropriately-generalized TT coordinates are denoted here by $(u,v,x^1, x^2)$, and are known in the literature as Rosen coordinates. They are related to their Brinkmann counterparts $(U,V,X^1, X^2)$ via
\begin{equation}
    u = U, \quad \bx = \bm{\xi}^{-1} (U) \bX, \quad v = V - \frac{1}{2} [\dot{\bm{\xi}}(U) \bm{\xi}^{-1}(U)]_{ij} X^i X^j,
    \label{BrinkmannToRosen}
\end{equation}
where $\xi_{ij}(U)$ is any nonsingular $2 \times 2$ (not necessarily symmetric) matrix satisfying
\begin{equation}
    \ddot{\bm{\xi}}(U) = \bH(U) \bm{\xi}(U), \qquad (\bm{\xi}^\intercal \dot{\bm{\xi}})_{[ij]} = 0.
    \label{Econstr}
\end{equation}
Any $\xi_{ij}$ with these properties is essentially a Jacobi propagator: Contracting it on the right with an arbitrary constant vector results in the nontrivial components of a solution\footnote{The geodesic deviation equation may be written as $\ddot{\bm{\psi}} = -\bm{R} \bm{\psi}$ \textit{in any spacetime}, where $\bm{\psi}$ denotes the four-dimensional deviation vector resolved into parallel-propagated tetrad components, $\bm{R}$ is a $4 \times 4$ matrix of similarly-decomposed Riemann components, and overdots represent ordinary derivatives with respect to an affine parameter \cite{Synge}.} 
to the geodesic deviation (or Jacobi) equation \cite{HarteCaustics}. Given any such propagator with the specified properties, the exact plane wave metric \eqref{Brinkmann} becomes
\begin{equation}
    \rmd s^2 = - 2 \rmd u \rmd v + \gamma_{ij}(u) \rmd x^i \rmd x^j,
    \label{Rosen}
\end{equation}
where 
\begin{equation}
    \gamma_{ij}(u) = \gamma_{(ij)} (u) \equiv \xi_{ki} (u) \xi_{kj} (u).
    \label{Hdef}
\end{equation}
The implied summation in this last equation is understood to be trivial in the sense that $\gamma_{ij} = \sum_k \xi_{ki} \xi_{kj}$, or equivalently $\bg = \bm{\xi}^\intercal \bm{\xi}$ in matrix notation. Regardless, the Rosen line element \eqref{Rosen} naturally splits into a longitudinal component $-2 \rmd u \rmd v$ and a transverse component $\gamma_{ij}(u) \rmd x^i \rmd x^j$. The longitudinal portion is flat and Lorentzian, while the transverse portion is associated with the Riemannian 2-metric $\gamma_{ij}(u)$. We refer to this 2-metric as the Rosen waveform, and note that it depends only on the phase $u$ of the gravitational wave. It is related to the Brinkmann waveform $H_{ij}(U)$ via \eqref{Econstr} and \eqref{Hdef}. 

That the Rosen coordinates are, as claimed, a type of geodesic normal coordinate system may be verified by noting that every fixed-$x^i$ 2-surface contains a timelike geodesic. Moreover, each such surface actually contains a 1-parameter family of timelike geodesics related by longitudinal boosts. It also contains a null geodesic along which $v = (\mbox{constant})$. These statements can be made more intuitive in terms of the quasi-Cartesian coordinates
\begin{equation}
    t = \frac{1}{\sqrt{2}} (v + u) , \qquad z = \frac{1}{\sqrt{2}} (v-u),
    \label{tzDef}
\end{equation}
defined by analogy with standard $\mbox{null $\rightarrow$ inertial}$ transformations in flat spacetime. The vector $\partial/\partial z$ is now a spatial projection of the wave propagation direction $\ell^a$. Additionally, any worldline which remains at fixed $(x^i,z)$ is a timelike geodesic. More generally, all curves satisfying $\rmd x^i/\rmd t = 0$ and $| \rmd z/\rmd t | = (\mbox{constant}) < 1$ are timelike geodesics. Worldlines constrained by $\rmd x^i/\rmd t = 0$, $|\rmd z/\rmd t| = 1$ are instead null geodesics. These statements imply that a large class of geodesics look trivial in Rosen coordinates. Geodesics which are not in this class can appear quite complicated, however. A generic timelike geodesic may be specified by choosing initial conditions for its transverse components at some fiducial phase $u = u_0$, as well as a constant $\lambda > 0$ which describes motion longitudinal to the wave. Defining the phase average of the inverse 2-metric by
\begin{equation}
    \langle \bm{\gamma}^{-1} \rangle \equiv \frac{1}{u- u_0} \int_{u_0}^u \bm{\gamma}^{-1} (\tau) d\tau,
    \label{gammaAv}
\end{equation}
an arbitrary timelike geodesic may be shown to have the coordinate parametrization
\begin{eqnarray}
    \fl \qquad \bx(u) &=& \bx(u_0) + (u-u_0) \langle \bm{\gamma}^{-1} \rangle \bm{\gamma} (u_0) \dot{\bx}(u_0),
    \label{GeodesicX}
    \\
    \fl \qquad v(u) &=& v(u_0) + (u - u_0) \left( \lambda^2 + \frac{1}{2} [\bm{\gamma} ( u_0 ) \langle \bm{\gamma}^{-1} \rangle \bm{\gamma} ( u_0 )]_{ij} \dot{x}^i( u_0 ) \dot{x}^j( u_0 )  \right).
    \label{GeodesicV}
\end{eqnarray}
Geodesics which remain at fixed $(x^i, z)$ satisfy $\dot{x}^i(u_0) = 0$ and $\lambda = 1$.

That the Rosen metric directly generalizes the TT-gauge plane wave \eqref{TTgauge} follows from applying \eqref{tzDef} to \eqref{Rosen}, which produces the exact line element $\rmd s^2 = - \rmd t^2 + \gamma_{ij}(u) \rmd x^i \rmd x^j + \rmd z^2$. Standard perturbative results are therefore recovered if
\begin{equation}
    \gamma_{ij}= \delta_{ij} + \epsilon h_{ij} + O(\epsilon^2)   
    \label{Hexpand}
\end{equation}
and $\tr \bm{h} = 0$. It is shown in Section \ref{Sect:perts} that such expansions do indeed arise when considering smooth 1-parameter families of plane wave spacetimes. While the vacuum Einstein equation is both linear and algebraic in terms of $H_{ij}$ [cf. \eqref{TrH}], it is a nonlinear differential equation when expressed in terms of $\gamma_{ij}$. This means that the $O(\epsilon^2)$ terms neglected in \eqref{Hexpand} are generically nontrivial. They may be viewed as corrections due to the higher-order Einstein equation, or alternatively as higher-order solutions to a family of geodesic equations.  

The second of these perspectives follows from the interpretation of the spatial Rosen coordinates as a lattice of timelike geodesics. The relative displacements of these geodesics are encoded in $\gamma_{ij}$. Remarkably, this matrix also encodes in a simple way many properties of the \textit{null} geodesics which are so central to the calculations of geometric optics. Plane wave spacetimes are sufficiently simple that most of the information required to characterize the propagation of light can i) be embedded in the Rosen-coordinate metric components, and ii) this coordinate choice is one for which considerable intuition and experience already exists (at least perturbatively). For these reasons, the remainder of this paper works almost exclusively in Rosen coordinates.

More precisely, it is assumed that a \textit{particular} Rosen coordinate system has been fixed. While the Brinkmann waveform $H_{ij}$ describing a particular plane wave is unique up to the relatively simple transformations \eqref{Htrans}, there exist many physically-equivalent Rosen waveforms $\gamma_{ij}$. Indeed, it is clear from \eqref{BrinkmannToRosen} that relations between these two types of waveforms are nonunique and also nonlocal; different solutions to \eqref{Econstr} are possible, and different resulting solutions for $\xi_{ij}$ generically define different $\gamma_{ij}$ via \eqref{Hdef}. This freedom corresponds to the ability to choose different collections of timelike geodesics as coordinate markers. It can be described more precisely as the set of all solutions to $\ddot{\bm{\xi}}= \bH \bm{\xi}$ for which $\bm{\xi}^\intercal \dot{\bm{\xi}}$ is symmetric (a constraint which is true everywhere if it is true anywhere), and for which $\bm{\gamma} = \bm{\xi}^\intercal \bm{\xi}$ is nonsingular, modulo those solutions which preserve $\bm{\xi}^\intercal \bm{\xi}$. Fixing a particular $H_{ij}$ while choosing a fiducial phase $u_0$, any two such solutions $\xi_{ij}$, $\xi_{ij}'$ to \eqref{Econstr} must be related by
\begin{eqnarray}
     \fl \bm{\xi}'(u) = \bm{\xi} (u) \left\{ \bm{\xi}^{-1} (u_0) \bm{\xi}' (u_0) + (u-u_0) \langle \bm{\gamma}^{-1} \rangle \left[ \bm{\xi}^\intercal (u_0) \dot{\bm{\xi}}' (u_0) - \dot{\bm{\xi}}^\intercal (u_0) \bm{\xi}'(u_0) \right] \right\}.
     \label{ETransform}
\end{eqnarray}
The complexity of this relation makes it clear that if one is presented with two Rosen waveforms $\gamma_{ij}$, $\gamma_{ij}'$, it not necessarily obvious by inspection whether or not they describe the same spacetime. It is nevertheless possible to check equivalence by computing some $\xi_{ij}$ and $\xi_{ij}'$ which generate the appropriate waveforms, and then determining if $\ddot{\bm{\xi}} \bm{\xi}^{-1}$ and $\ddot{\bm{\xi}}' \bm{\xi}'^{-1}$ can be related by the transformation \eqref{Htrans} [cf. \eqref{Econstr}].

Although it is rarely discussed, the same ambiguities regarding representations of gravitational waves arise even in the linearized theory. As a possible counterpoint, one might object that vacuum, first-order, TT-gauge metric perturbations on a flat background are known to be invariant with respect to first-order gauge transformations. Such statements depend, however, on global assumptions such as asymptotic falloff (cf. Section 2.3 of \cite{FlanaganHughes}). If a spacetime is taken to be a literal plane wave, there is no such falloff. If, more realistically, a plane wave is used only as a model intended to be valid in a finite region of spacetime, the true asymptotic boundary conditions are external to the modeled region and therefore unusable. From either perspective, the familiar uniqueness of TT-gauge metric perturbations is lost. If \eqref{Hexpand} holds, the transformations $h_{ij}(u) \rightarrow h_{ij}(u) + c_{ij} + u d_{ij}$ are easily shown to describe the same geometry through $O(\epsilon)$ for any constant trace-free matrices $c_{ij}$, $d_{ij}$. 

It is assumed here that a particular Rosen coordinate system has been chosen such that $\det \bm{\xi} > 0$ everywhere of interest. This ensures that a single coordinate patch can be used in all calculations, and also that the orientations of the $x^i$ and $X^i$ coordinates are identical. As with any geodesic-type coordinate system, gravitational focusing generically causes Rosen coordinate systems to break down on sufficiently large scales. Restricting to a single coordinate patch therefore implies an upper bound on the maximum distances over which the calculations described here can be applied. This bound is, for example, of order $(\epsilon \omega)^{-1}$ in the case of a monochromatic gravitational wave with angular frequency $\omega$ and strain amplitude $\epsilon$. Optical effects involving larger scales are discussed in \cite{HarteLensing}.

\subsection{Rosen metrics: Non-perturbative considerations}
\label{Sect:RosenDiscussion}

Suppose that a particular vacuum plane wave has been fixed by prescribing a trace-free matrix $H_{ij}(u)$. This is equivalent, via \eqref{Riemann}, to prescribing the wave's curvature. Rosen waveforms then follow from
\eqref{Econstr} and \eqref{Hdef}, implying that $\gamma_{ij}(u)$ is essentially the square of a matrix which describes the displacements of coupled parametric oscillators attached to ``springs'' whose squared natural frequencies\footnote{Einstein's equation requires that the eigenvalues of $H_{ij}$ have opposite signs, so the instantaneous frequencies in this analogy are not necessarily real.} are proportional to $H_{ij}(u)$. Similar equations arise throughout physics and engineering, and a variety of methods have therefore been developed to understand them.

A large class of physically-relevant waveforms can be classified as either ``burstlike''---having a large amplitude only for a short time---or ``continuous,'' implying approximate periodicity. The curvature associated with a burst might be further idealized as vanishing completely whenever $u$ lies outside a bounded connected region $\mathcal{C} \subset \mathbb{R}$. The spacetime can then be viewed as a curved region sandwiched between two flat connected regions $\mathcal{F}_\pm$. Such spacetimes are often described as ``sandwich waves.'' In either of the flat regions where $H_{ij} =0$, the general solution to \eqref{Econstr} is
\begin{equation}
    \bm{\xi} (u) = \bm{a}_{\mathcal{F}_\pm} + u \bm{b}_{\mathcal{F}_\pm},
    \label{xiFlat}
\end{equation}
The two matrices $\bm{a}_{\mathcal{F}_\pm}$, $\bm{b}_{\mathcal{F}_\pm}$ must be constant and must satisfy $(\bm{a}_{\mathcal{F}_\pm}^\intercal \bm{b}_{\mathcal{F}_\pm})_{[ij]}=0$ and $\det (\bm{a}_{\mathcal{F}_\pm} + u \bm{b}_{\mathcal{F}_\pm}) > 0$ everywhere of interest. 

The trivial case of a spacetime which is everywhere flat is conventionally described by a metric in which $\gamma_{ij} = \delta_{ij}$, implying that $\xi_{ij}$ must be a constant orthogonal matrix. This is not required, however. Regions of flat spacetime can also be covered by coordinate systems whose ``waveforms''---computed as squares of expressions like \eqref{xiFlat}---grow quadratically with $u$. Recalling \eqref{Brinkmann} and \eqref{BrinkmannToRosen}, this corresponds to using non-comoving geodesics as coordinate markers. While such possibilities are an uninteresting complication in spacetimes which are \textit{globally} flat, they can be unavoidable when considering bursts of gravitational waves. If coordinates are chosen such that $\gamma_{ij} = \delta_{ij}$ in one of a sandwich wave's flat regions, the 2-metric in the other flat region cannot be prescribed arbitrarily, but must instead be computed by integrating \eqref{Econstr} through the intervening burst of gravitational waves. Cases where resulting 2-metric is not equal to $\delta_{ij}$ physically correspond to gravitational wave bursts which impart permanent displacements (and possibly kicks) to initially-comoving test particles. This is known as the gravitational memory effect, and is discussed further in Sections \ref{Sect:Burst1} and \ref{Sect:Burst2} below.

Plane waves which cannot be modeled as bursts, but whose curvature is instead periodic in $u$, may be understood using different methods. In these cases, $\ddot{\bm{\xi}} = \bH \bm{\xi}$ corresponds to a set of coupled Hill-like equations. Although some methods exist for finding exact solutions to Hill equations \cite{ExactHill1, ExactHill2}, these are rather limited. General properties of periodic waves may instead be understood using Floquet theory \cite{HillBook}. It follows from this that even though $\xi_{ij}$ and $\gamma_{ij} = \xi_{ki} \xi_{kj}$ do not typically share the periodicity of $H_{ij}$, knowledge of $\xi_{ij}$ in only one oscillation period can be used to extend it everywhere. Although we shall not do so here, such methods can be used to greatly enlarge the validity of perturbative calculations. Another interesting consequence of Floquet theory is that it can be used to find which waves are ``unstable'' in the sense that they admit exponentially-growing $\xi_{ij}$. Such dramatic effects require very large gravitational wave amplitudes, however.

\section{Non-perturbative optical observables}
\label{Sect:Optics}

\begin{figure}
	\centering
	\begin{subfigure}[t]{.42\linewidth}
		\centering
		\includegraphics[width= \linewidth]{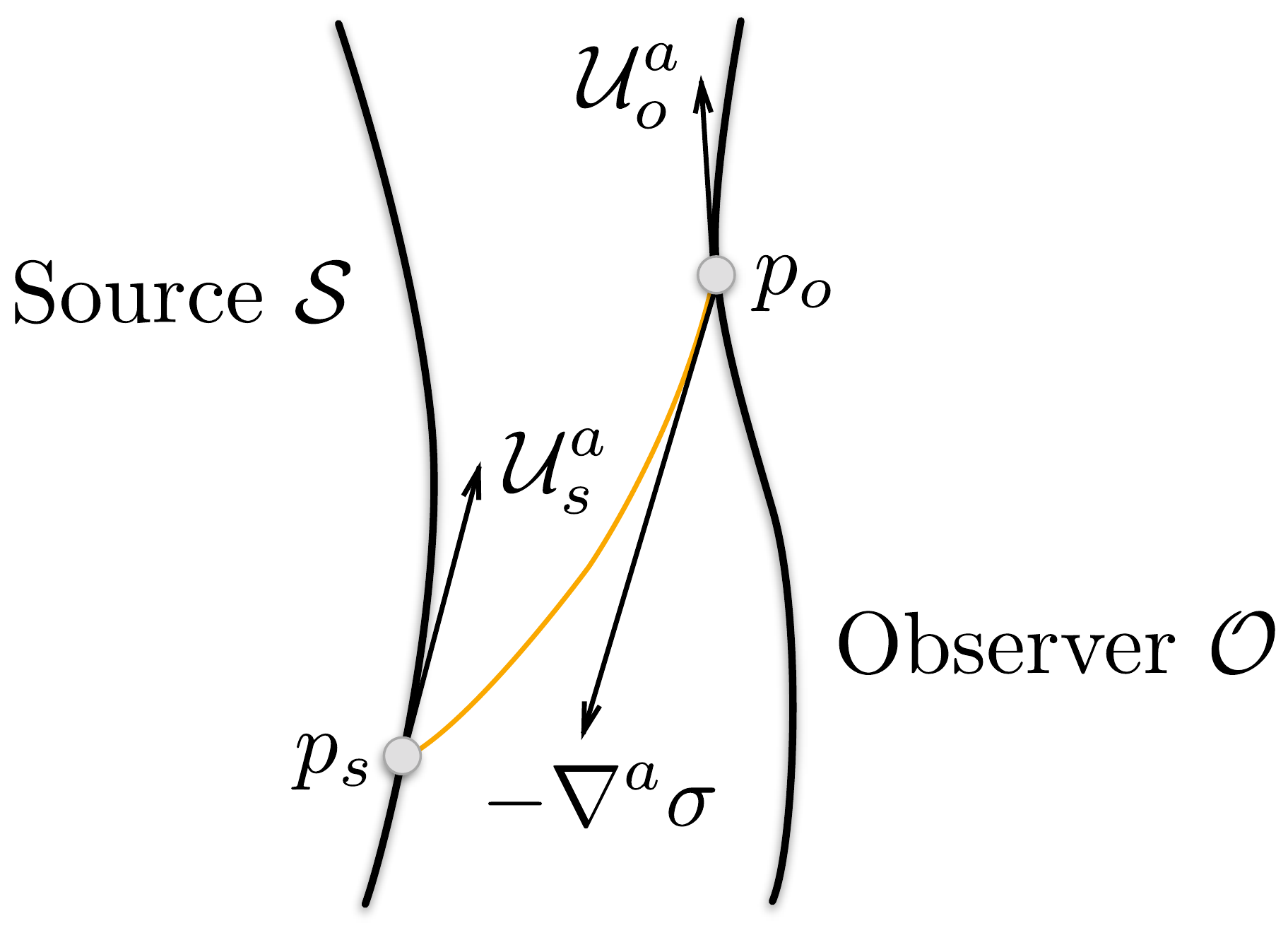}
		\caption{Diagram illustrating the observation of a pointlike source $\mathcal{S}$ by an observer $\mathcal{O}$. The observation event $p_o$ is connected to the emission event $p_s$ by a light ray whose tangent can be computed as the gradient of Synge's function $\sigma$. Source and observer 4-velocities are denoted by $\mathcal{U}^a_{o,s}$.}
    	\label{Fig:LensingGeom}
	\end{subfigure}
	~
	\begin{subfigure}[t]{.42\linewidth}
		\centering
		\includegraphics[width= .61\linewidth]{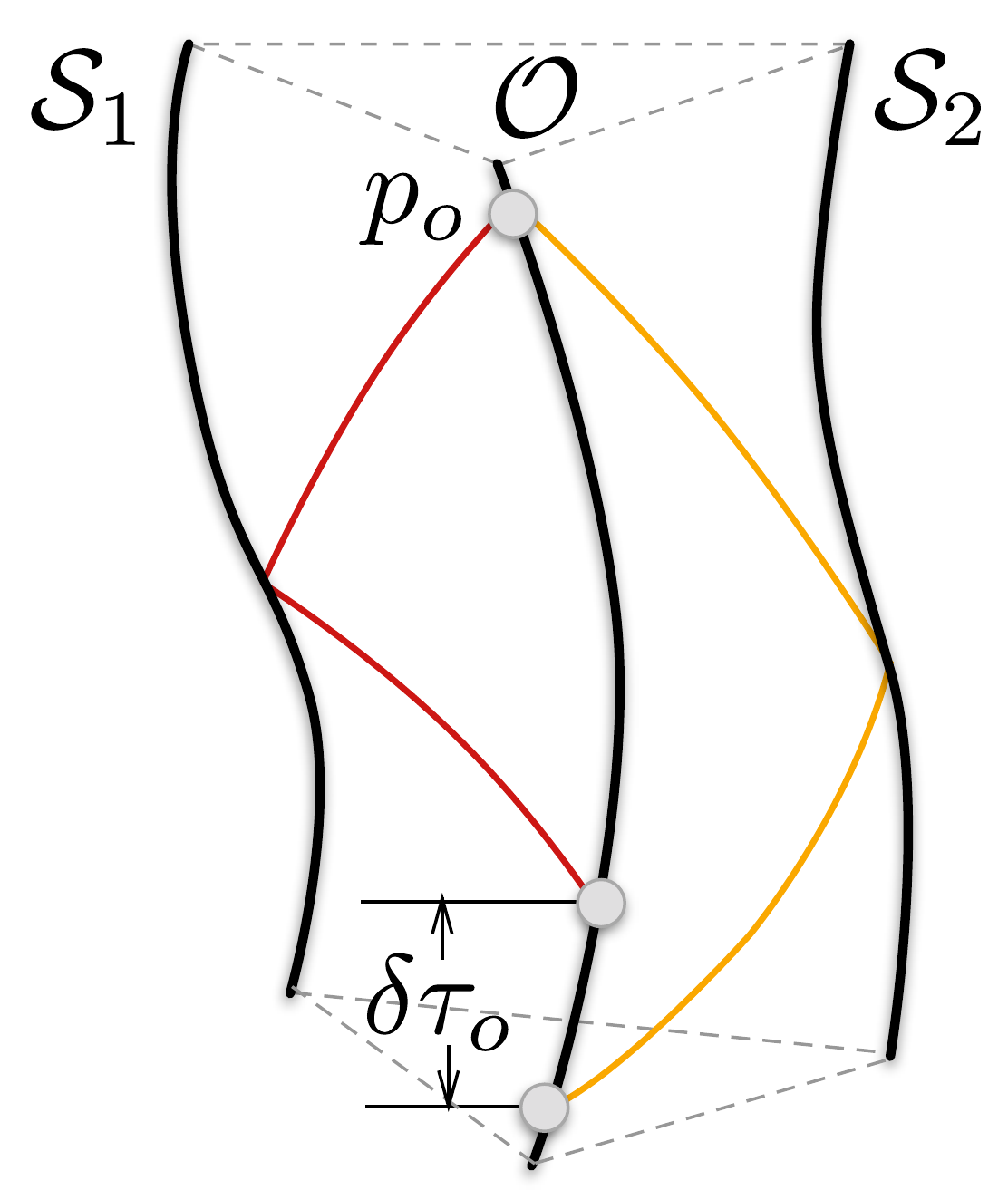}
		\caption{A two-arm interferometer with an observer $\mathcal{O}$ flanked by two mirrors $\mathcal{S}_{1,2}$. The relative phases of the light compared at $p_o$ determine the difference $\delta \tau_o$ in their emission times. This system may be viewed as a composite the one in part (a).}
		\label{Fig:Interferometer}
	\end{subfigure}
\end{figure}

How, at least in principle, might a gravitational wave be detected? Perhaps the most obvious observable associated with any spacetime is its curvature, and it follows from \eqref{Riemann} that knowledge of $R_{abc}{}^{d}$ for any gravitational plane wave immediately implies knowledge of its Brinkmann-type waveform. Local experiments may therefore be used to directly determine $H_{ij}(u)$ for all phases $u$ in which measurements are performed. Not all interesting observables are local, however. Pulsar timing techniques, for example, use radio bursts which are expected to travel through many gravitational wave cycles before reaching the earth. Observed properties of these bursts are therefore modified by a kind of integrated curvature which depends nontrivially on $H_{ij}$.

This section considers more generally those nonlocal observables which describe the appearance of a luminous source in the presence of an intervening gravitational plane wave. Exact equations are derived for the observed spectrum, sky location, angular size, and brightness of each such source. Some (though not all) of these observables have been considered previously in \cite{HarteLensing}, but in forms which were difficult to interpret and which could not be immediately compared to known perturbative results. Where overlap exists, the new equations obtained here are far simpler both to understand and to apply. Throughout, gravitational waveforms and polarization content are unconstrained, and light is assumed to follow the laws of geometric optics.

The basic system we consider is illustrated in Figure \ref{Fig:LensingGeom}. There, a source is abstracted to a timelike worldline $\mathcal{S}$ and its observer to another timelike worldline $\mathcal{O}$. Images then correspond to past-directed null geodesics from $\mathcal{O}$ to $\mathcal{S}$. Given any observation event $p_o \in \mathcal{O}$, exactly one image is assumed to exist in all cases considered here\footnote{There exist special source-observer configurations where pointlike sources appear as extended images. Plane wave spacetimes can also admit multiple discrete images on very large scales which cannot be described by the single Rosen coordinate patch assumed here \cite{HarteLensing, PerlickReview, FaraoniMult}.}. The remainder of this paper explores the properties of these images. More complicated optical problems (such as the interferometer illustrated in Figure \ref{Fig:Interferometer}) can typically be understood as multiple copies of the basic system shown in Figure \ref{Fig:LensingGeom}. Unless otherwise indicated, the results of this section place no restriction on the motions of the source or the observer.

\subsection{Time of flight}
\label{Sect:TOF}

Observations such as those illustrated in Figure \ref{Fig:LensingGeom} provide a natural mapping between ``observation times'' and ``emission times.'' More precisely, the light rays which connect $\mathcal{S}$ to $\mathcal{O}$ may be used to relate the proper time $\tau_o$ recorded by $\mathcal{O}$ at an observation event $p_o \in \mathcal{O}$ to the proper time $\tau_s$ recorded by $\mathcal{S}$ at the emission event $p_s\in \mathcal{S}$. Such relations play an important role in all optical calculations performed below. The redshift or blueshift can, for example, be computed from $\rmd \tau_s/\rmd \tau_o$. Additionally, the time difference $\delta \tau_o$ (or the phase shift) for the interferometer illustrated in Figure \ref{Fig:Interferometer} follows by successively applying the maps $\tau_s (\tau_o)$ applicable to each pair of optical elements. 

As a first step towards computing $\tau_s(\tau_o)$ for the simpler system in Figure \ref{Fig:LensingGeom}, consider instead the gravitational wave phase $u_s(u_o)$ at $p_s$ as a function of the gravitational wave phase $u_o$ at $p_o$.  Although the plane wave phase coordinate $u$ is null, the difference $u_o - u_s (u_o)$ may nevertheless be interpreted as a ``time of flight'' for a photon traveling from $\mathcal{S}$ to $\mathcal{O}$. This interpretation is supported by noting that i) the $u$ coordinate is unique up to a constant affine transformation, and ii) the restriction of $u$ to any timelike geodesic is, up to a positive constant, a proper time for that geodesic.

The phase relation $u_s(u_o)$ can be computed using Synge's function $\sigma(p,p')$. This is a two-point scalar which takes two events as arguments and returns one half of the squared geodesic distance between those events \cite{Synge, PoissonRev,Friedlander}. Plane wave spacetimes are one of the few examples where this function is known explicitly (although generic post-Minkowskian approximations are available \cite{TimeTransfer}), with the Brinkmann form appearing in, e.g., \cite{HarteCaustics} and the Rosen form in \cite{Friedlander}. Given two events $p$, $p'$ and their associated Rosen coordinates $(u,v,x^i)$, $(u',v',x'^i)$,
\begin{equation}
    \sigma(p,p') = -(u-u') (v-v') + \frac{1}{2} [ \langle \bm{\gamma}^{-1} \rangle^{-1} ]_{ij} (x-x')^i (x-x')^j,
    \label{Synge}
\end{equation}
where $\langle \bm{\gamma}^{-1} \rangle^{-1}$ denotes the matrix inverse of the average $\langle \bm{\gamma}^{-1} \rangle$ defined by \eqref{gammaAv}, except that this average is to be evaluated between $u'$ and $u$ instead of between $u_0$ and $u$. For the trivial case of a spacetime which is globally-flat, there exist coordinates where $\gamma_{ij} = \delta_{ij}$ and so $\sigma_\mathrm{flat}(p,p') = -(u-u') (v-v') + \frac{1}{2} | \bx - \bx'|^2 = \frac{1}{2} [ - (t-t')^2 + | \bx - \bx'|^2 + (z-z')^2]$. The general expression \eqref{Synge} for the geodesic distance in an arbitrary plane wave differs from this only via the replacement $\bm{I} \rightarrow \langle \bm{\gamma}^{-1} \rangle^{-1}$ in the transverse directions. 

The existence of the null geodesic which connects the observation and emission events illustrated in Figure \ref{Fig:LensingGeom} implies that
\begin{equation}
  \sigma(p_o, p_s) = \sigma(p_s, p_o) = 0,
  \label{SyngeVanish}
\end{equation}
so
\begin{equation}
    u_o - u_s = \frac{1}{2} (v_o-v_s)^{-1} [ \langle \bm{\gamma}^{-1} \rangle^{-1} ]_{ij} (x_o-x_s)^i (x_o-x_s)^j
    \label{uDelay}
\end{equation}
whenever $v_s \neq v_o$. Using the $t$ and $z$ coordinates defined by \eqref{tzDef} instead results in the equivalent
\begin{equation}
    t_o - t_s = \sqrt{ [\langle \bm{\gamma}^{-1} \rangle^{-1}]_{ij} (x_o - x_s)^i (x_o - x_s)^j + (z_o - z_s)^2}.
    \label{TimeDelay}
\end{equation}
Supplementing these equations with appropriate parametrizations for the source coordinates in terms of the gravitational wave phase allows explicit solutions to be obtained for $u_s(u_o)$ or $t_s(t_o)$. If the source and observer move on geodesics, the necessary parametrizations are given by \eqref{GeodesicX} and \eqref{GeodesicV}.

Once $u_s(u_o)$ is known, it may be used to relate proper times associated with the observation and emission events. The observer's proper time $\tau_o$ is related to the gravitational wave phase via
\begin{equation}
  \frac{\rmd \tau_o}{\rmd u} = - \frac{1}{\mathcal{U}^a_o \ell_a},
\end{equation}
where $\mathcal{U}^a_o$ denotes the observer's 4-velocity and $\ell^a = \partial/\partial v$ the gravitational wave's propagation direction. Noting that $\ell^a$ is Killing, this rate is a constant for geodesic observers. More generally, we parametrize it by
\begin{equation}
  \lambda_o \equiv - (\sqrt{2} \mathcal{U}^a_o \ell_a)^{-1} > 0,
  \label{kappaDef}
\end{equation}
which coincides with the $\lambda$ appearing in \eqref{GeodesicV} if the motion is geodesic. Regardless of acceleration, a (not necessarily constant) $\lambda_s$ can be defined analogously for the source, in which case proper times along $\mathcal{S}$ and $\mathcal{O}$ are related via
\begin{equation}
  \frac{\rmd \tau_s}{\rmd \tau_o} = (\lambda_s / \lambda_o ) \frac{ \rmd u_s }{ \rmd u_o } .
  \label{tauDot}
\end{equation}
If the source and the observer both move on geodesics, $\lambda_s/\lambda_o$ is a constant and this equation is trivially integrated to obtain $\tau_s(\tau_o)$ in terms of $u_s(u_o)$.

\subsection{Frequency shifts}

The derivative $\rmd \tau_s/ \rmd \tau_o$ appearing in \eqref{tauDot} is directly observable. If a physical process on $\mathcal{S}$ occurs with a characteristic frequency $\omega_s$---perhaps the frequency of a spectral line or the angular velocity of a pulsar---which is much larger than any frequencies associated with the gravitational wave or with the acceleration of  $\mathcal{S}$, such a process would be observed on $\mathcal{O}$ to occur at frequency $\omega_o = (\rmd \tau_s/ \rmd \tau_o) \omega_s$. This reduces to the familiar Doppler effect in flat spacetime, but more generally includes curvature corrections as well. One way to compute $\omega_o$ is to implicitly differentiate \eqref{uDelay} and then substitute the result into \eqref{tauDot}. Alternatively, note that the  gradient of Synge's function $\sigma(p_o,p_s)$ is tangent to the light ray which connects $p_s$ to $p_o$. It is also parallel-transported along that ray, implying that
\begin{equation}
  \frac{ \omega_o }{ \omega_s } = - \frac{ \mathcal{U}_o^a \nabla_a \sigma }{ \mathcal{U}^b_s \nabla_b \sigma }
  \label{RedshiftFund}
\end{equation}
in terms of the observer and source 4-velocities $\mathcal{U}^a_{o,s}$. The gradient in the numerator here is understood to be evaluated at $p_o$ while the gradient in the denominator is understood to instead be evaluated at $p_s$. It follows from \eqref{RedshiftFund} that $\omega_o = \omega_s$ if $\mathcal{U}^a_o$ is equal to $\mathcal{U}^a_s$ parallel-transported along the relevant light ray, a generalization of the flat-spacetime result that the Doppler effect vanishes for comoving objects. Such a condition is, however, physically unnatural in curved spacetimes; the frequency shift is generically nonzero.

Equation \eqref{RedshiftFund} may be directly evaluated using \eqref{Synge} and \eqref{kappaDef}. The result is conveniently expressed in terms of the 2-vectors
\begin{equation}
     \bm{k}_o \equiv \frac{\langle \bm{\gamma}^{-1} \rangle^{-1} (\bx_s - \bx_{o \rightarrow s}) }{\sqrt{2} \lambda_o ( u_o-u_s ) }, \qquad \bm{k}_s \equiv \frac{\langle \bm{\gamma}^{-1} \rangle^{-1} (\bx_{s \rightarrow o} - \bx_o ) }{\sqrt{2} \lambda_s ( u_o-u_s ) } ,
     \label{ko}
\end{equation}
where
\begin{equation}
      \fl \bx_{o \rightarrow s} \equiv \bx_o - (u_o - u_s) \langle \bm{\gamma}^{-1} \rangle \bm{\gamma}(u_o) \dot{\bx}_o, \quad  \bx_{s \rightarrow o} \equiv \bx_s + (u_o-u_s) \langle \bm{\gamma}^{-1} \rangle \bm{\gamma} (u_s) \dot{\bx}_s.
      \label{dXdef}
\end{equation}
All averages here are to be evaluated between $u_s$ and $u_o$, while, e.g., $\dot{\bx}_o$ denotes a $u$-derivative of the observer's transverse coordinates evaluated at $p_o$. Up to overall factors included for later convenience, $\bm{k}_o$ and $\bm{k}_s$ are essentially transverse separations between the source and the observer evaluated either at the emission phase $u_s$ or at the observation phase $u_o$. Causality does not allow any observation at $u = u_o$ to depend on properties of the source at that phase (except for special alignments where $u_s = u_o$), so the relevant observables instead involve an \textit{extrapolation} of the source's position from $u = u_s$ to $u = u_o$ performed using the geodesic which is tangent to $\mathcal{S}$ at $p_s$. Comparing \eqref{GeodesicX} and \eqref{dXdef}, $\bx_{s \rightarrow o}$ represents precisely this kind of ``osculating extrapolation.'' Similarly, $\bx_{o \rightarrow s}$ extrapolates the transverse location of the observer from $u = u_o$ to $u = u_s$ using the geodesic tangent to $\mathcal{O}$ at $p_o$. In terms of $\bm{k}_o$ and $\bm{k}_s$, the frequency ratio in the presence of an arbitrary plane wave is
\begin{equation}
  \frac{\omega_o}{\omega_s} = \frac{\rmd \tau_s}{\rmd \tau_o} = ( \lambda_o / \lambda_s ) \left( 1 + \frac{\bm{\gamma}^{-1}_{ij}(u_s) k^i_s k^j_s - \bm{\gamma}^{-1}_{ij}(u_o) k^i_o k^j_o}{ 1 + \bm{\gamma}_{ij}^{-1}(u_o) k^i_o k^j_o}\right)^{-1}.
  \label{RedshiftGeneral}
\end{equation}
This is valid for arbitrarily-moving sources and observers. It reduces in the flat limit $\gamma_{ij} \rightarrow \delta_{ij}$ to an expression for the ordinary Doppler shift. More generally, there is a sense in which the effects of relative motion and spacetime curvature can be disentangled by considering the time derivative of $\omega_o/\omega_s$ \cite{Helfer, Finn2, PTAgauge}. 

Equation \eqref{RedshiftGeneral} simplifies significantly if the source and observer are assumed to move on geodesics which remain at fixed spatial coordinates. When this occurs, $\lambda_o = \lambda_s = 1$ and
\begin{equation}
  \bm{k}_o = \bm{k}_s =\frac{ \langle \bm{\gamma}^{-1} \rangle^{-1} (\bx_s - \bx_o) }{ \sqrt{2} (u_o-u_s) } .
	\label{koSimp}
\end{equation}
The frequency shift then depends on the difference between inverse Rosen waveforms at the emission and observation events:
\begin{equation}
  \frac{\omega_o - \omega_s}{\omega_o} =  \frac{ [ \bm{\gamma}^{-1}_{ij}(u_o) - \bm{\gamma}^{-1}_{ij}(u_s) ] k^i_o k^j_o}{ 1 + \bm{\gamma}_{kl}^{-1}(u_o) k^k_o k^l_o}.
  \label{RedshiftStationary1}
\end{equation}
One deficiency with this formula is that $k^i_o$ has no immediate physical interpretation. It is closely related, however, to the observed position of the source. As explained in Section \ref{Sect:Position}, that position is naturally described in terms of a unit 3-vector $(\hat{\bm{k}}_\perp, \hat{k}_\|)$. Using \eqref{kHat} to relate this vector to $\bm{k}_o$, the frequency shift \eqref{RedshiftStationary1} can be rewritten as   
\begin{equation}
  \frac{\omega_o - \omega_s}{\omega_o} = \frac{1}{2}( 1 + \cos \theta )^{-1}  \hat{\bm{k}}_\perp^\intercal [ \bm{I} - \bm{\xi}(u_o) \bm{\gamma}^{-1} (u_s) \bm{\xi}^\intercal (u_o)] \hat{\bm{k}}_\perp ,
  \label{RedshiftStationary2}
\end{equation}
where $\theta$ denotes the observed angle between the source and the gravitational wave propagation direction [cf. \eqref{kToAngles}]. Additionally, $\xi_{ij}$ is a square root of $\gamma_{ij}$ in the sense of \eqref{Hdef}. Although such roots are not unique, a particular choice must be made to in order to define an observer rest frame with respect to which the components of $\hat{\bm{k}}_\perp$ are defined. The $\xi_{ij}$ appearing in \eqref{RedshiftStationary2} is the same as that used in \eqref{frameT} to construct this frame.

Although exact, \eqref{RedshiftStationary2} is very similar to the first-order perturbative formula typically used when discussing gravitational wave measurements via pulsar timing \cite{PTAreview, Detweiler} or Doppler velocimetry \cite{Estabrook, Doppler2, DopplerTracking}. In terms of the first-order metric perturbation $\epsilon h_{ij}$ which appears in \eqref{TTgauge}, it is well-known that
\begin{equation}
  \frac{ \omega_o - \omega_s }{ \omega_o } = \frac{\epsilon}{2} ( 1 + \cos \theta )^{-1} [ h_{ij}(u_s) - h_{ij}(u_o)] \hat{k}^i_\perp \hat{k}^j_\perp + O(\epsilon^2) .
  \label{Redshift1stOrder}
\end{equation}
The physical meanings of the $\theta$, $\hat{\bm{k}}_\perp$, and $u_s$ which appear here are unchanged in the exact result \eqref{RedshiftStationary2}, although their time-dependence is no longer trivial and relations to coordinate quantities are different. The metric difference $\epsilon [h_{ij}(u_s) - h_{ij}(u_o)]$ appearing in the first-order result is, however, generalized to 
\begin{equation}
  \delta_{ij} - \xi_{ik}(u_o) \gamma^{-1}_{kl} (u_s) \xi_{jl} (u_o) = [ \gamma^{-1}_{kl} (u_o) - \gamma^{-1}_{kl} (u_s) ] \xi_{ik} (u_o) \xi_{jl} (u_o).
  \label{gammaDiffEff}
\end{equation}
Direct approximations to \eqref{RedshiftStationary2} are discussed further in Section \ref{Sect:FrequencyPert}.

\subsection{Source locations}
\label{Sect:Position}

Gravitational waves may affect not only a source's apparent spectrum, but also its location on the sky. Such locations can be compactly described in terms of a 3-vector which resides in the observer's instantaneous rest frame. More precisely, consider an orthonormal triad $(\bm{e}^a_\perp, e^a_\|)$ which is orthogonal to the observer's 4-velocity $\mathcal{U}^a_o$. It is convenient to align this triad such that one of its components is locked to the direction of propagation associated with the background plane wave. Projecting the null propagation direction $\ell^a$ into the observer's rest frame and normalizing, the longitudinal frame vector is then
\begin{equation}
      e^a_\| \equiv \sqrt{2} \lambda_o ( g^{ab} + \mathcal{U}^a_o \mathcal{U}^b_o) \ell_b = \sqrt{2} \lambda_o \ell^a - \mathcal{U}^a_o . 
    \label{frameL}
\end{equation}
It is also convenient to define the remaining two frame vectors by
\begin{equation}
    \bm{e}^a_{\perp} \equiv \bm{\xi}^{-\intercal}(u_o) \left( \bg(u_o) \dot{\bx}_o \frac{\partial}{\partial v}  + \frac{\partial}{\partial \bx} \right),
    \label{frameT}
\end{equation}
where $\bm{\xi}$ is a particular matrix satisfying \eqref{Hdef}. It is easily verified that the resulting triad is indeed orthonormal and orthogonal to $\mathcal{U}^a_o$. It is also parallel-transported for geodesic observers. If $\mathcal{O}$ is not a geodesic, another frame---perhaps one which is Fermi-Walker transported---might be more natural. We nevertheless apply these definitions in all cases.
 
Recalling that the direction of the light ray connecting $p_o$ to $p_s$ is given by $- \nabla_a \sigma(p_o,p_s)$, its rest-frame components describe the observed location of $\mathcal{S}$. If the unit-normalized versions of these components are denoted by $(\hat{\bm{k}}^a_\perp, \hat{k}^a_\|)$, use of \eqref{Synge} results in
\begin{equation}
  \hat{\bm{k}}_\perp =  \frac{ 2  \bm{\xi}^{-\intercal}(u_o) \bm{k}_o }{ 1 + \bm{k}_o^\intercal \bg^{-1}(u_o) \bm{k}_o }, \qquad   \hat{k}_\| = \frac{ 1 - \bm{k}_o^\intercal \bg^{-1}(u_o) \bm{k}_o}{ 1 + \bm{k}_o^\intercal \bg^{-1}(u_o) \bm{k}_o },
  \label{kHat}
\end{equation}
where $\bm{k}_o$ is defined by \eqref{ko}. By construction, $|\hat{\bm{k}}_\perp|^2 + \hat{k}_\|^2 = 1$. These expressions hold for arbitrarily-moving sources and observers in arbitrary plane wave spacetimes.

It is often sufficient to consider only the transverse 2-vector $\hat{\bm{k}}_\perp$. This is proportional to $\bm{\xi}^{-\intercal}(u_o) \bm{k}_o$, the first factor of which takes into account that the transverse 2-metric $\bg(u_o) = \bm{\xi}^\intercal (u_o) \bm{\xi}(u_o)$ is not Euclidean at the observer's location. The vector $\bm{k}_o$ is more interesting, however. Recalling its definition, the observer's transverse location $\bx_o$ does not enter on its own, but instead appears via the osculating projection $\bx_{o \rightarrow s}$. Objects therefore appear to be not where they ``are,'' but where they ``would have been.'' This can be made more clear in terms of the distance measurements considered in Section \ref{Sect:Dist}. Using \eqref{theta} and the area distance $r_\mathrm{area}$ \eqref{dAng1} in \eqref{kHat},
\begin{equation}
    \fl \qquad \hat{\bm{k}}_\perp = \frac{1}{r_\mathrm{area}} \left(  \frac{ \det \langle \bm{\gamma}^{-1} \rangle }{ \sqrt{ \det \bm{\gamma}^{-1}(u_o) \bm{\gamma}^{-1}(u_s) } } \right)^{1/2} \bm{\xi}^{-\intercal}(u_o) \langle \bm{\gamma}^{-1} \rangle^{-1} (\bm{x}_s - \bm{x}_{o \rightarrow s}).
    \label{kHat2}
\end{equation}

It is sometimes convenient to parametrize source locations by angles instead of vectors. Let $(\theta,\phi)$ be polar coordinates constructed in the observer's sky such that $\theta$ corresponds to the apparent angle between the source and gravitational wave, while $\phi = 0$ coincides with the direction of $(\bm{e}^a_\perp)_1$. Then,
\begin{equation}
  \hat{\bm{k}}_\perp =
        \left(
        \begin{array}{cc}
          \cos \phi \sin \theta \\
          \sin \phi \sin \theta 
        \end{array}
        \right), \qquad 
        \hat{k} _\| = \cos \theta.
  \label{kToAngles}
\end{equation}
Combining this with \eqref{kHat} shows that the observed latitude $\theta$ explicitly satisfies
\begin{equation}
    \tan^2 (\theta/2) = \bm{k}_o^\intercal \bg^{-1}(u_o) \bm{k}_o .
    \label{theta}
\end{equation}

One interesting characteristic of position measurements is that they can be used to deduce the presence of a gravitational wave even when observing very nearby sources. While knowledge of $\hat{\bm{k}}_\perp$ at any one instant isn't particularly meaningful, its time evolution is. This evolution can be nontrivial no matter how close $\mathcal{S}$ happens to be to $\mathcal{O}$. It is evident from the presence of $\bg^{-1}(u_o) - \bg^{-1}(u_s)$ in \eqref{RedshiftStationary1} that useful timing measurements are instead restricted to source-observer separations over which significant differences can be expected in the gravitational waveform.

\subsection{Distances}
\label{Sect:Dist}

The next observable we consider is the area distance $r_\mathrm{area}$. If a source's angular size is resolvable and found to be equal to the small solid angle $\Omega$ in the observer's sky, there is a sense in which its physical area is $\Omega r_{\mathrm{area}}^2$ \cite{PerlickReview}. This is somewhat technical to derive, so we defer to the result obtained in \cite{HarteLensing} using Brinkmann coordinates. Translating that into Rosen coordinates while using \eqref{theta} results in
\begin{equation}
  r_\mathrm{area} = \frac{ \sqrt{2} \lambda_o (u_o-u_s) }{ 1 + \cos \theta}  \left(  \frac{ \det \langle \bm{\gamma}^{-1} \rangle }{ \sqrt{ \det \bm{\gamma}^{-1} (u_o) \bm{\gamma}^{-1} (u_s) }  } \right)^{1/2} .
  \label{dAng1}
\end{equation}
The first fraction here is essentially an affine distance to the source, while the second corrects this by taking into account the expansion of a thin bundle of light rays which converge on $p_o$. That latter correction is remarkably simple in terms of the Rosen waveform, depending only on the arithmetic and geometric averages of $\bg^{-1}$ between the source and the observer.

It can also be useful to describe a source in terms of its luminosity distance $r_\mathrm{lum}$, which is related to the area distance via
\begin{eqnarray}
    r_\mathrm{lum} = \left( \frac{ \rmd \tau_s }{ \rmd \tau_o} \right)^{-2} r_\mathrm{area}.
    \label{dLum}
\end{eqnarray}
One factor of $\rmd \tau_s/\rmd \tau_o$ occurs here due to the frequency shift experienced by light traveling from $\mathcal{S}$ to $\mathcal{O}$, while the second arises from considering bundles of light rays which converge on $p_s$ rather than $p_o$. An explicit formula for the luminosity distance follows immediately by substituting \eqref{RedshiftGeneral} and \eqref{dAng1} into \eqref{dLum}. For the special case where both sources and observers move on geodesics at fixed spatial coordinates, the frequency shift is given by \eqref{RedshiftStationary2} and hence
\begin{eqnarray}
    r_\mathrm{lum} &=& \frac{ \sqrt{2} (u_o-u_s) }{ 1 + \cos \theta}  \left(  \frac{ \det \langle \bm{\gamma}^{-1} \rangle }{ \sqrt{ \det \bm{\gamma}^{-1} (u_o) \bm{\gamma}^{-1} (u_s) }  } \right)^{1/2}	
    \nonumber
    \\
    && ~ \times \left[ 1 - \frac{1}{2} \left( \frac{\hat{\bm{k}}_\perp^\intercal [ \bm{I} - \bm{\xi}(u_o) \bm{\gamma}^{-1} (u_s) \bm{\xi}^\intercal (u_o)] \hat{\bm{k}}_\perp }{ 1 + \cos \theta} \right) \right]^2.
\end{eqnarray}

\section{Perturbative approach}
\label{Sect:Pert}

Equations \eqref{uDelay}, \eqref{RedshiftGeneral}, \eqref{kHat2}, \eqref{dAng1}, and \eqref{dLum} provide exact prescriptions for the time delays, frequency shifts, positions, and distances of generic sources in plane wave spacetimes. We now consider their perturbative expansions. This is done for two reasons: First, it provides a clear connection between the exact results derived here and the various approximations which have appeared in literature. Second, some physical implications of the optical formulae are more easily understood when written in an approximate form. In particular, the results of Section \ref{Sect:Optics} can be expanded beyond the lowest-order approximation which has typically been considered in the past, and doing so demonstrates that some higher-order corrections grow very rapidly with the source-observer distance. 

Two types of nonlinearity arise here. The first of these has its origin in the nonlinearity of Einstein's equation as applied to the Rosen-type plane wave metric \eqref{Rosen} and its associated waveform $\gamma_{ij}$. Perturbative expansions of this waveform are considered in Section \ref{Sect:perts}, and are sufficient to qualitatively understand that nonlinear effects can accumulate at large distances. The various gravitational wave observables are, however, nonlinear functionals of $\gamma_{ij}$. This is taken into account when deriving explicit second-order expansions for those observables in Section \ref{Sect:ObservablePerts}.

\subsection{Expanding the metric}
\label{Sect:perts}

As explained in Section \ref{Sect:Brinkmann}, gravitational plane waves in general relativity can be  trivially described in terms of the Brinkmann waveform $H_{ij}$ which appears in the line element \eqref{Brinkmann}. This directly determines the curvature, and is restricted by the vacuum Einstein equation only to be trace-free. The observables discussed in Section \ref{Sect:Optics} are not, however, written in terms of $H_{ij}$. They instead depend on the Rosen waveform $\gamma_{ij}$, which is related to $H_{ij}$ via the nonlinear and nonlocal expressions \eqref{Econstr} and \eqref{Hdef}. Although the vacuum Einstein equation could be applied to find a nonlinear differential equation for the Rosen waveform alone, we take the perspective that it is more natural to instead describe a plane wave in terms of $H_{ij}$ and then to derive $\gamma_{ij}$ from that. It is also convenient to view the perturbative expansion performed here not as an approximation for a single system which involves a small quantity $\epsilon$, but rather as an approximation for an entire \textit{family} of systems which are smoothly parametrized by $\epsilon$. Orders in perturbation theory then correspond to differentiations with respect to $\epsilon$ evaluated at $\epsilon = 0$.

The precise family of plane wave spacetimes considered here\footnote{It can sometimes be interesting to instead consider families of waves where the curvature depends smoothly on $\epsilon$, $u$ \textit{and} $u/\epsilon$. Any portions involving $u/\epsilon$ vary rapidly as $\epsilon \rightarrow 0$, thus evoking the concept of a ``high-frequency limit'' \cite{Isaacson,Burnett}. Considering such families is equivalent to a type a singular perturbation theory [as opposed to the regular perturbation associated with \eqref{Hfamily}]. See also the last paragraph of Section \ref{Sect:MonochromaticPert}.} is defined to be the set of Brinkmann line elements \eqref{Brinkmann} with
\begin{equation}
    H_{ij} (u;\epsilon) = \epsilon \mathfrak{H}_{ij}(u), \qquad \epsilon \geq 0.
    \label{Hfamily}
\end{equation}
The ``reference waveform'' $\mathfrak{H}_{ij}(u) = \partial_\epsilon H_{ij}(u;\epsilon)$ is assumed to be fixed and independent of $\epsilon$. Additionally supposing that $\tr \bm{\mathfrak{H}}(u) = 0$, it follows from \eqref{TrH} that each spacetime in this family is an exact solution to the vacuum Einstein equation. The parameter $\epsilon$ controls the amplitude of a wave's curvature, but not its polarization, frequency, or any other local properties. As expected from such an interpretation, different members of this family are physically distinct: Choosing any nonzero $\epsilon$ and any $\epsilon' \neq \epsilon$, the waveform $H_{ij} (u;\epsilon') = (\epsilon'/\epsilon) H_{ij} (u;\epsilon)$ cannot be transformed into $H_{ij} (u;\epsilon)$ using the gauge transformation \eqref{Htrans}. It is also clear that the wave disappears entirely entirely in the $\epsilon \rightarrow 0$ limit.

Given a trace-free $\mathfrak{H}_{ij}(u)$, an associated Rosen waveform $\gamma_{ij}(u;\epsilon)$ may be found by first constructing a matrix $\xi_{ij}(u;\epsilon)$ with the expansion
\begin{equation}
    \bm{\xi}(u;\epsilon) = \bm{\xi}_{(0)} (u) + \ldots + \epsilon^n \bm{\xi}_{(n)}(u) + \ldots .
    \label{Eorders}
\end{equation}
Applying \eqref{Econstr},
\begin{equation}
     \ddot{\bm{\xi}}_{(0)} = 0, \qquad \ddot{\bm{\xi}}_{(n)} = \bm{\mathfrak{H}} \bm{\xi}_{(n-1)}
    \label{EconstrPert}
\end{equation}
for all $n \geq 1$. Iteratively solving these equations then results in an approximation for $\xi_{ij} (u;\epsilon)$. Using it together with \eqref{Hdef} produces a family of Rosen waveforms $\gamma_{ij}(u;\epsilon)$ with the perturbative expansion coefficients
\begin{equation}
    \bm{\gamma}_{(n)}(u) \equiv \frac{1}{n!} \left. \partial_{\epsilon}^n \bm{\gamma}(u;\epsilon) \right|_{\epsilon=0} = \sum_{p+q=n} \bm{\xi}_{(p)}^\intercal (u) \bm{\xi}_{(q)} (u).
    \label{gammaIterate}
\end{equation}
Recalling \eqref{TTgauge}, we use the more standard notation $\bm{h}(u)$ interchangeably with $\bg_{(1)}(u)$ for the first-order metric perturbation.

As already emphasized in Section \ref{Sect:BrinkToRosen}, different Rosen waveforms---corresponding to different initial conditions for \eqref{Econstr}---can be used to describe the same physics. Ambiguities of this kind are easily resolved for the sandwich waves described in Section \ref{Sect:RosenDiscussion}, in which case it is natural to fix a particular waveform by demanding that $\gamma_{ij} = \xi_{ij} = \delta_{ij}$ in one of the locally-flat regions. More generally, however, there does not appear to be any ``preferred'' choice.

Our approach is to fix a convenient solution for $\xi_{ij}$, and then to note that all other possibilities are related via \eqref{ETransform}. Letting this particular solution satisfy the non-perturbative initial condition
\begin{equation}
    \xi_{ij}(u_0;\epsilon) = \delta_{ij}, \qquad \dot{\xi}_{ij} (u_0;\epsilon) = 0
    \label{XiData}
\end{equation}
for some constant $u_0$, it follows from \eqref{EconstrPert} that $\bm{\xi}_{(0)} = \bm{I}$ and
\begin{equation}
    \bm{\xi}_{(n)}(u) = \int_{u_0}^u \! \rmd u_1 \int_{u_0}^{u_1} \!\! \rmd u_2 \bm{\mathfrak{H}}(u_2) \bm{\xi}_{(n-1)}(u_2) 
    \label{xiIterate}
\end{equation}
and for all $n \geq 1$. Hence,
\begin{eqnarray}
    \tr \bm{\xi}_{(n)} = \bm{\xi}_{(n)}^\intercal - \bm{\xi}_{(n)} = 0 \qquad & \mbox{($n$ odd)},
\end{eqnarray}
and
\begin{eqnarray}
    \bm{\xi}_{(n)} + \bm{\xi}_{(n)}^\intercal = (\tr \bm{\xi}_{(n)}) \bm{I} \qquad & \mbox{($n$ even)}.
\end{eqnarray}
Combining these constraints with \eqref{gammaIterate} shows that each $\bg_{(n)}$ generated by a $\bm{\xi}$ which satisfies \eqref{XiData} is symmetric and that
\begin{equation}
    \tr \bm{\gamma}_{(n)} = 0 \quad (\mbox{$n$ odd}) , \qquad \bm{\gamma}_{(n)} = \frac{1}{2} ( \tr  \bm{\gamma}_{(n)} ) \bm{I} \quad (\mbox{$n$ even}).
\end{equation}
The familiar tracelessness of the first-order TT-gauge perturbation $\bm{h} = \bg_{(1)}$ therefore generalizes to all odd-order metric perturbations. It does not generalize to even orders; those expansion coefficients are instead ``pure trace.''

Through second order, these results imply that the particular Rosen waveform defined by \eqref{XiData} is explicitly
\begin{eqnarray}
   \fl \bm{\gamma}(u;\epsilon)  &=& \bm{I} + 2 \epsilon \int_{u_0}^{u} \! \rmd u_1 \int_{u_0}^{u_1} \!\! \rmd u_2 \bm{\mathfrak{H}}(u_2) + \epsilon^2 \tr \left[ \frac{1}{2} \left( \int_{u_0}^{u} \! \rmd u_1 \int_{u_0}^{u_1} \!\! \rmd u_2 \bm{\mathfrak{H}}(u_2) \right)^2 \right.
    \nonumber
    \\
     \fl && \left. ~ + \int_{u_0}^{u} \! \rmd u_1 \int_{u_0}^{u_1} \!\! \rmd u_2 \int_{u_0}^{u_2} \!\! \rmd u_3 \int_{u_0}^{u_3} \!\! \rmd u_4 \bm{\mathfrak{H}}(u_2) \bm{\mathfrak{H}}(u_4)  \right] \bm{I} + O(\epsilon^3) .
    \label{H2ndOrder}
\end{eqnarray}
The second line of this equation generically grows with $u$, which can be seen more clearly if the waveform is rewritten in terms of the first-order perturbation $h_{ij}$ instead of $\mathfrak{H}_{ij} = \frac{1}{2} \ddot{h}_{ij}$. Making this replacement and integrating by parts results in
\begin{eqnarray}
   \fl \bm{\gamma}(u;\epsilon)  &=& \bm{I} + \epsilon \bm{h}(u) + \frac{1}{4} \epsilon^2 \tr \left[ \bm{h}^2(u) - \int_{u_0}^{u} \! \rmd u_1 \int_{u_0}^{u_1} \!\! \rmd u_2 \dot{\bm{h}}^2(u_2)  \right] \bm{I} + O(\epsilon^3) .
    \label{H2ndOrderAlt}
\end{eqnarray}
The trace of $\dot{\bm{h}}^2$ cannot be negative, so the double integral here---and therefore the second-order metric perturbation as a whole---typically grows with $u$. Indeed, there exists an $\epsilon$-dependent lengthscale beyond which the second-order term outpaces the first. We now consider two examples which demonstrate this explicitly.

\subsubsection{Gravitational wave bursts with memory}
\label{Sect:Burst1}

Consider a sandwich wave as described in Section \ref{Sect:RosenDiscussion}. More specifically, consider an $\epsilon$-dependent family of waves \eqref{Hfamily} with $\mathfrak{H}_{ij}(u)$ zero everywhere except in a small region $u \in (-\delta, 0)$ inside of which the curvature oscillates several times. Choosing $u_0 < - \delta$ in \eqref{XiData} guarantees that the metric before the burst is trivial: $\gamma_{ij}(u;\epsilon) = \delta_{ij}$. The metric after the burst must, however, be generated by the square of \eqref{xiFlat}, where $\bm{a}_{\mathcal{F}_+} (\epsilon)$ and $\bm{b}_{\mathcal{F}_+} (\epsilon)$ are constant matrices determined by the details of the curved region. If $\bm{a}_{\mathcal{F}_+}(\epsilon) \neq \bm{I}$ or $\bm{b}_{\mathcal{F}_+}(\epsilon) \neq 0$, the gravitational wave burst is said to exhibit memory \cite{MemoryOriginal1,MemoryOriginal2} in the sense that it permanently displaces initially-comoving pairs of freely-falling test particles. It follows from \eqref{xiIterate} that the first and second-order perturbations are
\begin{eqnarray}
	(\bm{a}_{\mathcal{F}_+})_{(1)} = \frac{1}{2} \bm{h}(0) = \int_{-\delta}^0 \! \rmd u_1 \int_{-\delta}^{u_1} \!\! \rmd u_2 \bm{\mathfrak{H}}(u_2), 
	\\
	(\bm{b}_{\mathcal{F}_+})_{(1)} = \frac{1}{2} \dot{\bm{h}}(0) = \int_{-\delta}^0 \! \rmd u_1 \bm{\mathfrak{H}}(u_1),
\end{eqnarray}
and
\begin{eqnarray}
    (\bm{a}_{\mathcal{F}_+})_{(2)} = \frac{1}{4} \tr \left[ (\bm{a}_{\mathcal{F}_+})_{(1)}^2 - \frac{1}{2} \int_{-\delta}^0 \!\! \rmd u_1 \int_{-\delta}^{u_1} \!\! \rmd u_2 \dot{\bm{h}}^2(u_2) \right] \bm{I},
    \label{a2}
    \\
      (\bm{b}_{\mathcal{F}_+})_{(2)} = \frac{1}{2}  \tr \left[ (\bm{a}_{\mathcal{F}_+})_{(1)} (\bm{b}_{\mathcal{F}_+})_{(1)} - \frac{1}{4} \int_{-\delta}^0 \rmd u_1 \dot{\bm{h}}^2 (u_1) \right] \bm{I}    .
      \label{b2}
\end{eqnarray}

Although gravitational wave bursts with memory can arise from many physical scenarios, classic discussions \cite{MemoryOriginal2, Garfinkle} consider those waves which are emitted either from the scattering of multiple masses or the explosion of one mass into multiple unbound components. Applying the quadrupole formula in such cases suggests that $(\bm{b}_{\mathcal{F}_+})_{(1)} = 0$.\footnote{The condition $(\bm{b}_{\mathcal{F}_+})_{(1)} = 0$ implies that through first order, pairs of test particles which are comoving before the burst are also comoving after the burst; only their displacements might be permanently affected at $O(\epsilon)$. Some cases where an additional ``velocity memory'' arises are discussed in \cite{Polnarev}. See also \cite{Garfinkle, Bessonov} for an electromagnetic memory effect where nontrivial velocity changes occur even at lowest order in simple scattering problems.} Assuming this, a coordinate rotation may always be performed together with a rescaling of $\epsilon$ such that
\begin{equation}
    (\bm{a}_{\mathcal{F}_+})_{(1)} = \frac{1}{2} 
        \left( \begin{array}{cc}
            1    &    0\\
            0    &    -1
        \end{array} \right) .
\end{equation}
Now consider second order effects. It follows from \eqref{b2} that even though $\bm{b}_{\mathcal{F}_+}$ vanishes at first order, it \textit{cannot} vanish at second order. Indeed, $\tr (\bm{b}_{\mathcal{F}_+})_{(2)} < 0$ in these cases. Applying \eqref{Hdef} and \eqref{xiFlat}, the second-order waveform is explicitly
\begin{eqnarray}
  \bg(u;\epsilon) &=& \bm{I} + \epsilon \left\{ 
          \left( \begin{array}{cc}
            1    &    0\\
            0    &    -1
        \end{array} \right) + \frac{1}{4} \epsilon \left[ 1 + 4 \tr (\bm{a}_{\mathcal{F}_+})_{(2)} \right] \bm{I} \right\} 
        \nonumber
        \\
        && ~+ \epsilon^2 u \big[ \tr  (\bm{b}_{\mathcal{F}_+})_{(2)}\big] \bm{I} + O(\epsilon^3)
        \label{gammaApproxMem}
\end{eqnarray}
when $u_o > 0$. If the first-order waveform inside a burst is schematically of the form $h(u) \sim \cos \omega u$ and oscillates $N \sim \omega \delta$ times, it follows from \eqref{b2} that $\tr (\bm{b}_{\mathcal{F}_+})_{(2)} \sim - N \omega$. The second-order metric perturbation therefore becomes comparable to the first-order perturbation when
\begin{equation}
    \omega u \sim (\epsilon N  )^{-1} \gg 1.
    \label{epsilonBoundBurst}
\end{equation}
Of course, $\bg_{(1)}$ and $\bg_{(2)}$ are qualitatively different when this occurs. One is trace-free while the other is a pure trace. On even larger scales where $\omega u \sim (\epsilon^2 N )^{-1}$, the determinant of the Rosen waveform goes to zero. Recalling the line element \eqref{Rosen}, the metric itself becomes singular on this scale. Such effects are not an artifact of the perturbative expansion, but instead signal the breakdown of the Rosen coordinate system due to the gravitational focusing of nearby geodesics.

\subsubsection{Monochromatic waves}
\label{Sect:MonochromaticPert}

Another instructive example is provided by a linearly-polarized gravitational wave with constant amplitude. Let this wave be monochromatic with angular frequency $\omega$ in the sense that it can be described by the family \eqref{Hfamily} with
\begin{equation}
    \bm{\mathfrak{H}} (u) = \frac{\omega^2 }{2}
        \left( \begin{array}{cc}
            1    &    0\\
            0    &    -1 
        \end{array} \right)
    \cos \omega u .
    \label{HfrakMono}
\end{equation}
It follows from \eqref{Econstr} and \eqref{Hfamily} that if $\omega = 1$, the $2,2$-component of $\xi_{ij}(u;\epsilon)$ satisfies the Mathieu equation \eqref{Mathieu}. [The equation for general $\omega$ differs from this only via the rescaling $u \rightarrow \omega u$, resulting in $\ddot{\xi}_{22}(u;\epsilon) + \frac{1}{2} \epsilon \omega^2 \xi_{22}(u;\epsilon) = 0$.] We now show that important features of the metric perturbations associated with monochromatic gravitational waves are captured by the approximate Mathieu solution \eqref{MathieuExpand}. 

\begin{figure}
	\centering
	\includegraphics[width=.7\linewidth]{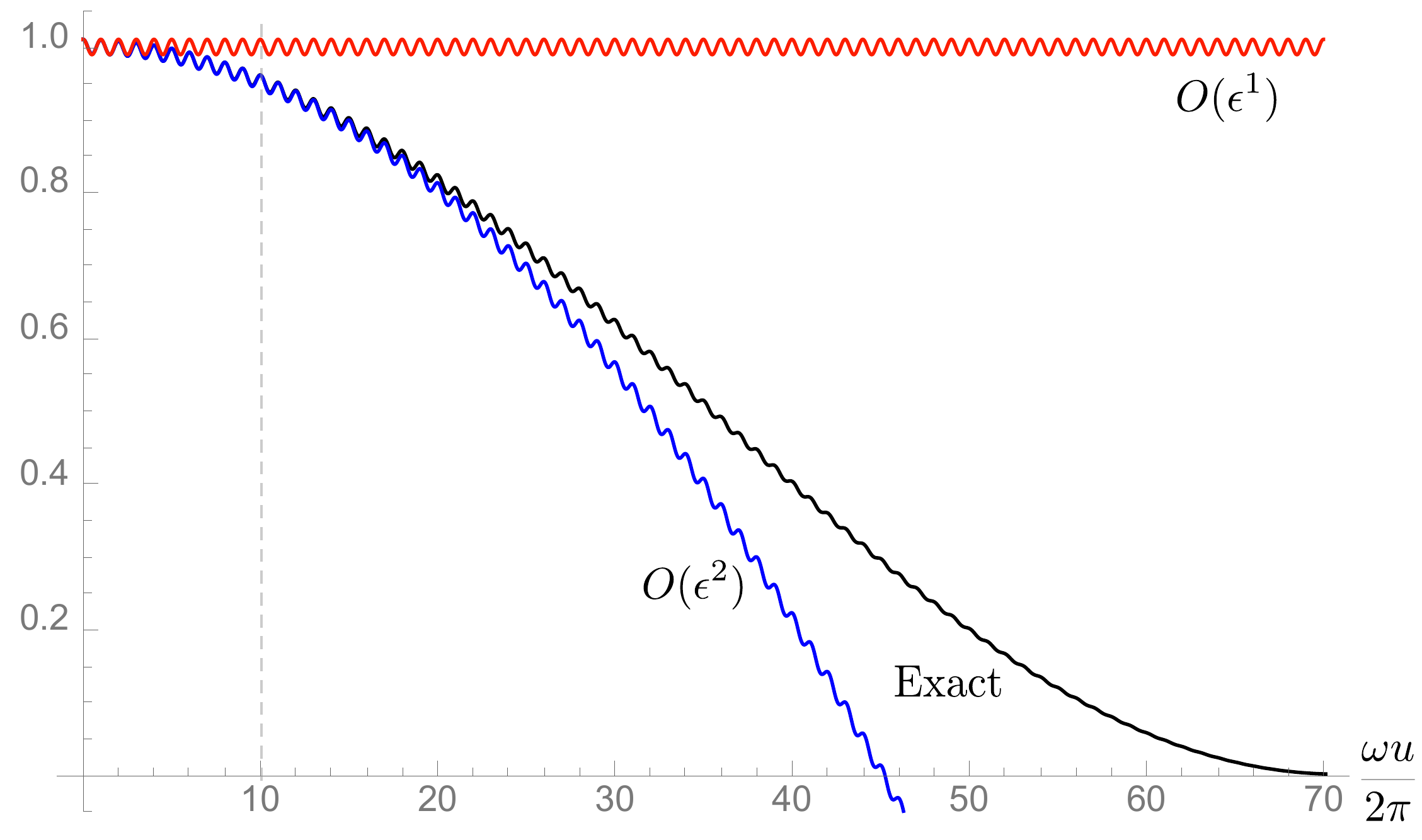}
	\caption{Plot of $\gamma_{22}(u)$ for a monochromatic gravitational wave with  $\mathfrak{H}_{ij}(u)$ given by \eqref{HfrakMono} and $\epsilon = 10^{-2}$. An exact solution is shown together with its first and second-order approximations. Initial conditions for all three curves are matched at $u=0$. The $O(\epsilon)$ approximation is poor after $\epsilon^{-1/2} = 10$ curvature oscillations, although the $O(\epsilon^2)$ approximation still works well there. A coordinate singularity appears in fewer than $\epsilon^{-1} = 100$ oscillations.}
	\label{Fig:MetricApprox}
\end{figure}

Although it is possible to express all components of $\gamma_{ij} (u;\epsilon)$ exactly in terms of Mathieu functions, consider instead the perturbative expansion described above. This first requires imposing initial conditions for $\xi_{ij}(u; \epsilon)$. Unfortunately, there is no natural phase at which to apply \eqref{XiData}. Indeed, all choices are more complicated than necessary. Setting $u_0 = 0$ for definiteness, \eqref{gammaIterate} and \eqref{xiIterate}  imply that $h_{ij}$ involves an oscillating term as well as a constant offset. Eliminating this offset by modifying the initial condition at $O(\epsilon)$ recovers the expected first-order perturbation
\begin{equation}
    \bm{h} (u) = -
        \left( \begin{array}{cc}
            1    &    0\\
            0    &    -1
        \end{array} \right)
    \cos \omega u .
    \label{hMono}
\end{equation}
The second-order perturbation is then
\begin{equation}
      \bm{\gamma}_{(2)} (u) = \frac{1}{16} [ 1 - 2 (\omega u)^2 +3 \cos 2 \omega u ] \bm{I} + 2 [\bm{\xi}_{(2)} (0) + u \dot{\bm{\xi}}_{(2)} (0)],
      \label{gamma2Cont}
\end{equation}
where the final two terms allow for $O(\epsilon^2)$ modifications to the initial condition \eqref{XiData}. It is clear, however, that no matter what these modifications are, the quadratic growth when $|\omega u| \gg 1$ cannot be eliminated. This is a second-order effect which becomes comparable to first order effects when
\begin{equation}
    |\omega u| \sim \epsilon^{-1/2} \gg 1.
    \label{epsilonBoundCont}
\end{equation}
It can be understood intuitively as a version of \eqref{epsilonBoundBurst} where the number of gravitational wave oscillations $N$ isn't fixed, but is instead of order $\omega u$. For a given wave amplitude $\epsilon$, it follows that nonlinear effects tend to be more important for approximately-monochromatic waves than for bursts. Also note that as suggested by \eqref{gamma2Cont}, Rosen coordinates break down when $|\omega u| \sim \epsilon^{-1} \gg \epsilon^{-1/2}$. This is illustrated explicitly in Figure \ref{Fig:MetricApprox} for the special case $\epsilon = 10^{-2}$.

The large nonlinear effects considered thus far do not oscillate. Monochromatic waves where $\epsilon \ll 1$ first experience relatively-large oscillatory corrections at $O(\epsilon^3)$. Assuming initial conditions for $\bm{\xi}_{(1)}$ such that \eqref{hMono} holds, the large-$u$ portions of $\bg_{(2)}$ and $\bg_{(3)}$ can be viewed as conformal corrections to the first-order waveform: Using an ellipsis to denote omitted terms at second and third order which grow more slowly than $u^2$,
\begin{eqnarray}
    \fl \qquad \quad \bg(u;\epsilon) &=& \left[  1 - \frac{1}{8}  (\epsilon \omega u)^2 \right] \left[ \bm{I} - \epsilon
        \left( \begin{array}{cc}
            1    &    0\\
            0    &    -1
        \end{array} \right)
    \cos \omega u \right] + \ldots + O (\epsilon^4).
    \label{gamma3Cont}
\end{eqnarray}

The growing nonlinearities discussed here can be interpreted in some ways as manifestations of an averaging effect which occurs more explicitly in ``high-frequency'' perturbations where families of waveforms $H_{ij}(u;\epsilon)$ are considered which are smooth in $u/\epsilon$ and $\epsilon$, but not in $u$ and $\epsilon$. In that context, the zeroth-order metric is no longer Minkowski, but instead satisfies the non-vacuum Einstein equation with an effective stress-energy tensor sourced by the gravitational wave \cite{Isaacson, Burnett}. Indeed, this stress-energy tensor is controlled by an average of the same ``energy density'' $\dot{\bm{h}}^2$ which is responsible for the second-order metric \eqref{H2ndOrderAlt} in an ordinary perturbative expansion. The high-frequency viewpoint is not pursued here firstly because its formulation requires choosing one small parameter which simultaneously controls both wave amplitudes and distances. This somewhat limits flexibility and clarity for the questions considered here. More importantly, different observables can react very differently when considering non-smooth families of spacetimes. The usual high-frequency formulations in general relativity are designed to compute metrics which are limits of solutions to Einstein's equation. It is not necessarily true, however, that limits of optical observables can be easily described in terms of these same metrics (or even that their limits exist at all).

\subsection{Gravitational wave observables: perturbative expressions}
\label{Sect:ObservablePerts}

Perturbative expansions for the various observables computed in Section \ref{Sect:Optics} may now be found for families of plane waves described by \eqref{Hfamily}. Completely general expressions would be quite complicated, so we restrict attention only to those cases where the observer $\mathcal{O}$ remains fixed at the spatial coordinates $(\bx_o, z_o)$. Also suppose that the source $\mathcal{S}$ remains fixed at $(\bx_s, z_s)$. As guaranteed by the construction of the Rosen coordinate system described in Section \ref{Sect:BrinkToRosen}, all such worldlines are timelike geodesics. They are not preserved, however, by changes in the Rosen waveform; coordinate transformations \eqref{ETransform} which modify $\gamma_{ij}$ also impart initially-stationary geodesics with nonzero coordinate velocities. Our assumption that particular sources and observers remain at fixed spatial coordinates is therefore coupled to the choice of a particular Rosen waveform. Although no detailed prescription is used here, we do assume that
	\begin{equation}
  		\fl \bm{\xi}_{(0)} = \bg_{(0)} = \bm{I}, \qquad \bm{\xi}_{(1)} = \bm{\xi}_{(1)}^\intercal = \frac{1}{2} \bg_{(1)},  \qquad \tr \bg_{(1)} = 0, \qquad \bg_{(2)} \propto \bm{I}. 
  	\label{RosenAssume}
	\end{equation}
	These equations are implied by the initial condition \eqref{XiData}, although they also hold more generally. 

\subsubsection{Zeroth order}
\label{Sect:ZerothOrder}

Our first step is to compute all observables in the flat-spacetime limit $\epsilon \rightarrow 0$. Denoting orders in perturbation theory by analogy with \eqref{Eorders}, the zeroth-order observables are
\begin{eqnarray}
  r_\mathrm{area}^{(0)} = r_\mathrm{lum}^{(0)} = r \equiv \sqrt{ | \bm{x}_s - \bm{x}_o|^2 + (z_s-z_o)^2 },
  \\
  \omega_o^{(0)} - \omega_s =  0,
  \label{Redshift0}
  \\
  \hat{\bm{k}}_\perp^{(0)} = (\bx_s - \bx_o)/r ,
  \label{k0}
  \\
  \cos \theta_{(0)} = (z_s - z_o)/r.
\end{eqnarray}
As expected, both the area and luminosity distances reduce to the coordinate distance $r$ at this order. Frequency shifts also vanish, and a source's apparent location in the observer's sky is given by the expected function of its coordinate position. Somewhat less obviously, \eqref{uDelay} implies that the zeroth-order emission phase $u_s^{(0)}$ is
\begin{equation}
      u_s^{(0)} =u_o -  \frac{r}{ \sqrt{2} } (1+ \cos \theta_{(0)})
      \label{us0}
\end{equation}
in terms of the observation phase $u_o$. 

All higher-order perturbative expressions obtained below depend in various ways on these zeroth-order quantities. Consistently retaining ``$(0)$'' subscripts or superscripts on each of them would considerably increase clutter, so this notation is suppressed when no confusion should arise.

\subsubsection{Time of flight}

The first-order perturbation $u_s^{(1)} \equiv \partial_\epsilon u_s |_{\epsilon = 0}$ to the zeroth-order emission phase \eqref{us0} is easily found by differentiating  \eqref{uDelay} with respect to $\epsilon$ while holding $u_o$ fixed. This results in
\begin{equation}
  \frac{ u_s^{(1)} }{ r } = - \frac{1}{2 \sqrt{2}}  \langle h_{ij} \rangle \hat{k}^i_\perp \hat{k}^j_\perp,
  \label{us1}
\end{equation}
where $\hat{\bm{k}}_\perp$ is understood to refer to the zeroth-order source direction $\hat{\bm{k}}_\perp^{(0)}$. Similarly, $\langle h_{ij} \rangle$ denotes an average of the first-order metric perturbation between $u_s^{(0)}$ and $u_o$. Decomposing $h_{ij}$ into its $+$ and $\times$ components using \eqref{hPlane}, the emission phase perturbation can alternatively be written as
	\begin{equation}
		\frac{ u_s^{(1)} }{ r }  = - \frac{1}{2 \sqrt{2}}  \left[ \langle h_+ \rangle \cos 2 \phi + \langle h_\times \rangle \sin 2 \phi \right] \sin^2 \theta.
		\label{us1Alt}
	\end{equation}
This vanishes for all sources which are aligned ($\theta_{(0)} = 0$) or anti-aligned ($\theta_{(0)} = \pi$) with the gravitational wave. If that wave is linearly-polarized, coordinates may be chosen such that $h_\times = 0$, implying that $u_s^{(1)}$ also vanishes for any source which is nominally located on one of the four meridians
	\begin{equation}
  		\phi_{(0)} = (1 + 2 n) \pi/4, \qquad n = 0,1,2,3.
  		\label{phiVanish}
	\end{equation}	
Regardless of polarization, the averages appearing in \eqref{us1Alt} are typically small if the waveform is approximately oscillatory and there are many oscillations between the source and the observer. These averages can be important, however, when there is significant gravitational memory.

The second-order perturbation $u_s^{(2)} \equiv \frac{1}{2} \partial_\epsilon^2 u_s|_{\epsilon = 0}$ to the emission phase is somewhat more complicated. Differentiating \eqref{uDelay} a second time with respect to $\epsilon$ shows that
\begin{eqnarray}
  \frac{ u_s^{(2)} }{ r } &=& - \frac{ 1 }{ 4 \sqrt{2} } \tr \left[ \langle \bg_{(2)} \rangle  + \langle \bm{h} \rangle^2 - \langle \bm{h}^2 \rangle  \right] \sin^2 \theta 
  \nonumber
  \\
  && ~ + \frac{ u_s^{(1)} }{2 r}   \left( \frac{ h_{ij}(u_s) - \langle h_{ij} \rangle   }{1 + \cos \theta } - \frac{1}{2} \langle h_{ij} \rangle \right) \hat{k}^i_{\perp} \hat{k}_{\perp}^j .
  \label{us2}
\end{eqnarray}
Unlike its first-order analog, $u_s^{(2)}$ does not tend to zero for oscillatory, memory-free waves at large distances. Part of it is also independent of $\phi_{(0)}$.

\subsubsection{Frequency shifts}
\label{Sect:FrequencyPert}

A perturbative expansion for the relative frequency shift $(\omega_o - \omega_s)/\omega_s$ induced by a gravitational plane wave may be found by differentiating the exact result \eqref{RedshiftStationary2} with respect to $\epsilon$ while holding $\omega_s$ fixed. At first order, 
\begin{equation}
    \frac{ \omega_o^{(1)} }{ \omega_s }  = \frac{1}{2} ( 1 + \cos \theta)^{-1} [ h_{ij}(u_s) - h_{ij} (u_o) ] \hat{k}^i_\perp \hat{k}^j_\perp,
  \label{Redshift1}
\end{equation}
which is immediately seen to be equivalent to the well-known \cite{PTAreview, Detweiler} approximation \eqref{Redshift1stOrder} when combined with \eqref{Redshift0}. In terms of $h_+$ and $h_\times$, it is more explicitly
\begin{equation}
  \fl \quad \frac{ \omega_o^{(1)} }{ \omega_s }  = \frac{1}{2} ( 1 - \cos \theta) \big\{ [ h_+(u_s) - h_+ (u_o) ] \cos 2 \phi + [ h_\times(u_s) - h_\times (u_o) ] \sin 2 \phi \big\}.
	  \label{Redshift1Alt}
\end{equation}
Equation \eqref{us0} implies that $u_s^{(0)} \rightarrow u_o$ as $\theta_{(0)} \rightarrow \pi$, so $\omega_o^{(1)} = 0$ for any sources which are aligned or anti-aligned with the background gravitational wave. If the wave is linearly-polarized and coordinates are chosen such that $h_\times = 0$, first-order frequency shifts also vanish when $\phi_{(0)}$ satisfies \eqref{phiVanish}.

Regardless of polarization, differentiating \eqref{RedshiftStationary2} again with respect to $\epsilon$ results in the second-order correction
\begin{eqnarray}
    \fl \frac{ \omega_o^{(2)} }{ \omega_s } = \frac{1}{4} ( 1 - \cos \theta) \tr \left\{ \bg_{(2)} (u_s) - \bg_{(2)} (u_o) + [\bm{h}(u_s) - \bm{h}(u_o)] [ 2 \langle \bm{h} \rangle - \bm{h} (u_s ) - \bm{h}(u_o)] \right\}
    \nonumber
    \\
    \fl \quad ~ + \frac{1}{2} (1+ \cos \theta)^{-1} \left[ (\omega_o^{(1)}/\omega_s) [ h_{ij}(u_s) - (3 + \cos \theta) \langle h_{ij} \rangle ] + u_s^{(1)} \dot{h}_{ij}(u_s) \right] \hat{k}^i_\perp \hat{k}^j_\perp.
    \label{Redshift2}
\end{eqnarray}
The second-order metric difference $\bg_{(2)}(u_s^{(0)}) - \bg_{(2)} (u_o)$ which appears here is directly analogous to the first-order difference $\bm{h} (u_s^{(0)}) - \bm{h}(u_o)$ found in $\omega_o^{(1)}/\omega_s$. Also note that the final term involving $\dot{\bm{h}}(u_s)$ may be interpreted as a straightforward correction to the first-order expression obtained by the replacement $\bm{h}(u_s^{(0)}) \rightarrow \bm{h}(u_s^{(0)} + \epsilon u_s^{(1)} + \ldots)$. Overall, $\omega_o^{(2)}$ and $\omega_o^{(1)}$ both vanish when $\theta_{(0)} \rightarrow 0$ or $\theta_{(0)} \rightarrow \pi$.

\subsubsection{Source locations}

The apparent location of a source is governed via \eqref{kToAngles} by the unit vector $(\hat{\bm{k}}_\perp , \hat{k}_\|)$ introduced in Section \ref{Sect:Position}. Differentiating the exact expression \eqref{kHat} for $\hat{\bm{k}}_\perp$ with respect to $\epsilon$ shows that first-order astrometric effects associated with a gravitational plane wave follow from
\begin{equation}
     \fl \quad \qquad \hat{\bm{k}}^{(1)}_\perp = \big[ \langle \bm{h} \rangle - \frac{1}{2} \bm{h} (u_o) \big] \hat{\bm{k}}_\perp + \frac{1}{2} \left[ \left( \frac{ h_{ij}(u_o) - \langle h_{ij} \rangle  }{1 + \cos \theta } -  \langle h_{ij} \rangle \right) \hat{k}^i_\perp \hat{k}^j_\perp \right] \hat{\bm{k}}_\perp ,
     \label{k1}
\end{equation}
a result which has also been derived (using different methods) in \cite{BookFlanagan}. One of its consequences is that the angle $\theta$ between the source and the gravitational wave is perturbed by
\begin{equation}
    \theta_{(1)} = \frac{1}{2}  \Big\{ \big[ \langle h_{ij} \rangle - h_{ij}(u_o) \big] + \langle h_{ij} \rangle \cos \theta \Big\} \hat{k}^i_\perp \hat{k}^j_\perp \csc \theta .
    \label{theta1}
\end{equation}
If a wave is linearly-polarized with $h_\times = 0$, sources which nominally lie on the meridians \eqref{phiVanish} experience vanishing latitudinal motion. These sources can, however, appear to rotate slightly around the gravitational wave propagation direction: $\theta_{(1)} = 0$ but $\phi_{(1)} = \pm [ \langle h_+ \rangle - \frac{1}{2} h_+ (u_o) ]$.

Regardless of polarization, two special cases of \eqref{k1} may be understood immediately. The first of these supposes that $\langle h_{ij} \rangle$ can be neglected, and may be thought of as a ``large-$r$'' limit in the presence of memory-free waves. Applying it,
\begin{eqnarray}
     \fl \qquad \qquad \hat{\bm{k}}^{(1)}_\perp \rightarrow \frac{1}{2} \Big\{ (1-\cos \theta) \big[ h_+(u_o) \cos 2 \phi + h_\times(u_o) \sin 2 \phi \big] \bm{I} -\bm{h} (u_o) \Big\} \hat{\bm{k}}_\perp .
     \label{k1Far}
\end{eqnarray}
Gravitational waves can also produce nontrivial astrometric effects in a ``small-$r$'' limit where $\langle h_{ij} \rangle \rightarrow h_{ij}(u_o)$ and
\begin{eqnarray}
     \hat{\bm{k}}^{(1)}_\perp \rightarrow \frac{1}{2} \Big\{ \bm{h} (u_o) - \sin^2 \theta \big[ h_+(u_o) \cos 2 \phi + h_\times(u_o) \sin 2 \phi \big] \bm{I} \Big\} \hat{\bm{k}}_\perp .
     \label{k1Close}
\end{eqnarray}
In either of these cases, first-order position perturbations depend on the waveform only at the observer (where $u = u_o$). The angular dependence of this effect is nevertheless distinct for near and distant sources.

If a gravitational wave is linearly-polarized, the matrices which multiply $\hat{\bm{k}}^{(0)}_\perp$ in \eqref{k1Far} and \eqref{k1Close} depend on $u_o$ only via an overall scaling. Sources in either limit therefore appear to move coherently along straight lines in the presence of linearly-polarized gravitational waves. The orientations and relative amplitudes of these lines  depend, however, on each source's nominal location $(\theta_{(0)}, \phi_{(0)})$. More complicated apparent motions can arise in either the small-$r$ or large-$r$ limits for gravitational waves which are not linearly-polarized, and even for linearly-polarized waves when $\langle h_{ij} \rangle$ is neither negligible nor approximately equal to $h_{ij} (u_o)$.

Astrometric effects are significantly more complicated at higher orders. For brevity, we therefore present second-order corrections only for the latitude $\theta$. Expanding \eqref{theta},
\begin{eqnarray}
    \fl  \theta_{(2)} &=& \frac{1}{4} \tr \bigg\{ \! \left[ \langle \bg_{(2)} \rangle - \bg_{(2)} (u_o) \right] + \langle \bg_{(2)} \rangle \cos \theta  + (1+ \cos \theta)  \left[ \langle \bm{h} \rangle^2 - \langle \bm{h}^2 \rangle \right] 
    \nonumber
    \\
    \fl && ~  + \left[ \langle \bm{h} \rangle - \bm{h}(u_o) \right]^2 \bigg\} \sin \theta + \bigg\{ \frac{ u_s^{(1)} }{\sqrt{2} r} \bigg[  \frac{ \langle h_{ij} \rangle - 2 h_{ij}(u_o) }{ 1 + \cos \theta}  + \frac{1}{2} ( 4 + \cos \theta) \langle h_{ij} \rangle 
    \nonumber
    \\
    \fl && ~ - h_{ij}(u_s) \bigg] \hat{k}^i_\perp \hat{k}^j_\perp - \frac{1}{2} (2 - \cos \theta) \theta^2_{(1)} \bigg\} \csc \theta.
    \label{theta2}
\end{eqnarray}
All terms involving $\bg_{(2)}$ in this equation are easily seen to be direct generalizations of the first-order metric perturbations appearing in $\theta_{(1)}$ [cf. \eqref{theta1}].

\subsubsection{Distances}

The remaining observables considered here are the area distance $r_\mathrm{area}$ and the luminosity distance $r_\mathrm{lum}$. First-order perturbations for these quantities may be found by expanding the exact expressions \eqref{dAng1} and \eqref{dLum}, which results in
\begin{eqnarray}
    \frac{ r_\mathrm{area}^{(1)} }{ r } = \frac{1}{2} \left(  \langle h_{ij} \rangle + \frac{ \langle h_{ij} \rangle - h_{ij}(u_o) }{ 1 + \cos \theta}  \right) \hat{k}^i_\perp \hat{k}^j_\perp ,
    \label{rArea1}
\end{eqnarray}
and
\begin{equation}
    \frac{ r^{(1)}_\mathrm{lum} }{ r } = \frac{ r^{(1)}_\mathrm{area} }{ r } + (1+\cos \theta)^{-1}   [  h_{ij}(u_o) - h_{ij} (u_s) ] \hat{k}^i_\perp \hat{k}^j_\perp  . 
    \label{rLum1}
\end{equation}
The second of these results describes how an object's apparent brightness is affected by a gravitational wave. It has sometimes been stated in the literature that gravity first affects brightnesses only at second order in general relativity (e.g., \cite{Zipoy, FaraoniScint}). A typical argument appeals to the Raychaudhuri equation, which can be used to show that the expansion of a null congruence is unperturbed through first order \textit{for objects at a fixed affine distance}. This is misleading, however. Gravitational waves (and more general geometries) do affect affine distances at $O(\epsilon)$. This and the first-order time dilation both contribute nontrivial first-order perturbations to $r_\mathrm{area}$ and $r_\mathrm{lum}$.

Second-order distance perturbations must take into account changes in a source's affine distance, time dilation, and, unlike in the first-order case, the gravitational focusing of null congruences. All of these effects are taken into account automatically by differentiating \eqref{dAng1} twice with respect to $\epsilon$:
\begin{eqnarray}
	\fl  \frac{ r_\mathrm{area}^{(2)} }{ r } && = \frac{1}{4} \tr \Big\{ \bg_{(2)} (u_o) \cos \theta + \bg_{(2)}(u_s) - (1+\cos \theta) \cos \theta \left[ \langle \bg_{(2)} \rangle + \langle \bm{h} \rangle^2 - \langle \bm{h}^2 \rangle \right]
	\nonumber
	\\
	\fl && ~ + (1-\cos \theta) [ \langle \bm{h} \rangle - \bm{h}(u_o)]^2 - \frac{1}{2} [ \bm{h}^2 (u_o) + \bm{h}^2 (u_s) - 2 \langle \bm{h} \rangle^2 ] \Big\} + \left( \frac{ u_s^{(1)}/\sqrt{2} r }{ 1 + \cos \theta } \right)
	\nonumber
	\\
	\fl && ~ \times \left[ \frac{ 2 \langle h_{ij} \rangle - h_{ij}(u_o) - h_{ij}(u_s) }{ 1 + \cos \theta } + \frac{1}{2} (3 + \cos \theta) \langle h_{ij} \rangle - h_{ij}(u_s) \right] \hat{k}^i_\perp \hat{k}^j_\perp .
	\label{rArea2}
\end{eqnarray}
The second-order perturbation to the luminosity distance follows from this, \eqref{Redshift1}, \eqref{Redshift2}, and \eqref{rArea1} via
\begin{equation}
  \frac{ r_\mathrm{lum}^{(2)} }{ r } = \frac{ r_\mathrm{area}^{(2)} }{ r } -2 \big( \omega_o^{(1)} / \omega_s \big) \frac{r_\mathrm{area}^{(1)} }{r} + 3 \big( \omega_o^{(1)} / \omega_s  \big)^2 - 2 \big( \omega_o^{(2)} / \omega_s  \big) .
  \label{rLum2}
\end{equation}

\subsection{Summary of second-order effects}
\label{Sect:2ndOrderSummary}

Most terms in the second-order expressions derived above are relatively uninteresting in the sense that their magnitudes are comparable to squares of typical first-order magnitudes. If these latter $O(\epsilon)$ effects are only marginally detectible, their squares are hopelessly small. More interesting are the second-order terms which can acquire large numerical coefficients. As motivated in Section \ref{Sect:perts}, $\bg_{(2)}$ can be large, thus implying that \eqref{us2} and \eqref{theta2} simplify to
\begin{eqnarray}
	u_s^{(2)} = - \frac{r}{4 \sqrt{2}} \tr \langle \bg_{(2)} \rangle \sin^2 \theta + \ldots , 
	\label{us2Large}
	\\
	\theta_{(2)} = \frac{1}{4} \tr \left\{ \left[ \langle \bg_{(2)} \rangle - \bg_{(2)} (u_o) \right] + \langle \bg_{(2)} \rangle \cos \theta \right\} \sin \theta + \ldots ,
	\label{theta2Large}
\end{eqnarray}
at large distances. These are identical to the expressions which would be obtained if the first-order expressions \eqref{us1} and \eqref{theta1} for $u_s^{(1)}$ and $\theta_{(1)}$ were applied with the substitution $\bm{h} \rightarrow \epsilon \bg_{(2)}$, providing a sense in which the potentially-significant contributions to $u_s$ and $\theta$ through second order mimic first-order effects with an ``effective first-order metric'' $\bm{h} + \epsilon \bg_{(2)}$. This effective metric has a nonzero trace, and therefore mimics a third type of gravitational wave polarization---sometimes referred to as a ``breathing mode'' \cite{AltGrav}---which introduces a distinct $\phi$-dependence into various observables. It must be emphasized, however, that this ``extra polarization'' is only an effective phenomenon. It is entirely determined by the ordinary $+$ and $\times$ polarization states, and therefore does not represent a physically-independent degree of freedom. This is also an effect which arises only when considering nonlocal observables, and not in, e.g., direct local measurements of $R_{abc}{}^{d}$.

Although the notion of an effective metric applied to first-order expressions describes all potentially-large second-order contributions to $u_s$ and $\theta$, it cannot do so for all interesting terms 
\begin{eqnarray}
\fl \frac{ \omega_o^{(2)} }{ \omega_s } = \frac{1}{4} (1 - \cos \theta) \tr [ \bg_{(2)} (u_s) - \bg_{(2)} (u_o) ]  + \frac{1}{2} \big( u_s^{(1)} / r \big) \left( \frac{ r \dot{h}_{ij}(u_s) \hat{k}^i_\perp \hat{k}^j_\perp }{ 1 + \cos \theta } \right) + \ldots  
\label{Redshift2Large}
\end{eqnarray}
associated with the second-order frequency perturbation \eqref{Redshift2}. The trace terms here are indeed those which would be obtained by adding an appropriate correction to the metric perturbation appearing in $\omega_o^{(1)}/\omega_s$. The remaining portion of the second-order frequency perturbation is different, however. It arises from the wave-induced perturbation to the emission phase, and can be important when $h_{ij}$ varies significantly over scales of order $\epsilon u_s^{(1)}$. Somewhat more precisely, use of \eqref{us1} shows that the $\dot{h}_{ij}$ term in \eqref{Redshift2Large} can be large if $\langle h_{ij} \rangle$ is significant \textit{and} there is a sense in which $r \dot{h}_{ij} \gg h_{ij}$. This occurs if, e.g., a gravitational wave simultaneously possesses both high and low frequency components. See Section \ref{Sect:BurstAndCont}.

The effective metric concept fails completely to describe the second-order perturbations to the area and luminosity distances. Using \eqref{rArea2} and \eqref{rLum2}, these perturbations are dominated by
\begin{equation}
  \fl \qquad \qquad \frac{ r_\mathrm{area}^{(2)} }{ r } = \frac{1}{4} \tr \left[ \bg_{(2)}(u_o) \cos \theta + \bg_{(2)} (u_s) - (1 + \cos \theta) \cos \theta \langle \bg_{(2)} \rangle \right] + \ldots 
  \label{rArea2Large}
\end{equation}
and
\begin{eqnarray}
  \frac{ r_\mathrm{lum}^{(2)} }{ r } = \frac{1}{4} \tr \big[ (2 - \cos \theta) \bg_{(2)}(u_o) + (2 \cos \theta - 1) \bg_{(2)} (u_s) 
  \nonumber
  \\
  ~ - (1 + \cos \theta) \cos \theta \langle \bg_{(2)} \rangle \big]   - \big( u_s^{(1)}/r \big) \left( \frac{ r \dot{h}_{ij}(u_s) \hat{k}^i_\perp \hat{k}^j_\perp }{ 1 + \cos \theta } \right) + \ldots 
  \label{rLum2Large}
\end{eqnarray}
at large distances, which could not have been guessed from the first-order expressions \eqref{rArea1} and \eqref{rLum1}. This is because the gravitational focusing of neighboring null geodesics is essential to both of these expressions, but has no first-order analog in vacuum general relativity.

\section{Gravitational wave examples}
\label{Sect:Example}

We now consider the optical effects of three types of gravitational waves in order  to illustrate some physical consequences of the perturbative expressions derived in Section \ref{Sect:Pert}. The first of these examples involves a gravitational wave which is both monochromatic and linearly-polarized. Next under discussion is a fast burst with nontrivial memory. Lastly, we consider a superposition of these two possibilities. 

\subsection{Optics in a monochromatic wave}
\label{Sect:ExMono}

Perhaps the simplest physically-interesting gravitational wave is a linearly-polarized example whose curvature is monochromatic in the sense described in Section \ref{Sect:MonochromaticPert}. Specifically, consider a family \eqref{Hfamily} of waves with curvatures given by \eqref{HfrakMono}. Also suppose that Rosen coordinates have been chosen such that $\bg_{(1)} = \bm{h}$ is given by \eqref{hMono} and $\bg_{(0)} = \bm{I}$. 

The first-order frequency shift for freely-falling sources and observers embedded in such a wave and remaining at fixed spatial coordinates follows from \eqref{Redshift1Alt}:
\begin{equation}
  \omega_o^{(1)}/ \omega_s  = \frac{1}{2} ( \cos \omega u_o - \cos \omega u_s ) ( 1 - \cos \theta) \cos 2 \phi  .
\end{equation}
This generically oscillates as $u_o$ is varied, but vanishes for any sources with $\phi_{(0)}$ satisfying \eqref{phiVanish}. Other observables depend on the average waveform $\langle h_{ij} \rangle$ between the emission and observation events. Using \eqref{us0} to define the zeroth-order estimate 
\begin{equation}
  N \equiv \frac{ \omega r }{ 2^{3/2} \pi} ( 1 + \cos \theta) 
  \label{NDef}
\end{equation}
for the number of gravitational wave cycles between these events, $\langle h_{ij} \rangle \sim N^{-1}$ over large distances where $N \gg 1$. It follows that averages can be ignored at first order in the large-distance limit. Equation \eqref{us1} then implies that $u_s^{(1)} \rightarrow 0$. The first-order position change is nontrivial, however, and may be described by
\begin{equation}
  \theta_{(1)} \rightarrow \frac{1}{2} \cos \omega u_o \sin \theta \cos 2 \phi , \qquad \phi_{(1)} \rightarrow - \frac{1}{2} \cos \omega u_o \sin 2 \phi .
\end{equation}
It also follows from \eqref{rArea1} and \eqref{rLum1} that the first-order perturbations to a source's apparent distance are
\begin{eqnarray}
  r_\mathrm{area}^{(1)}/r \rightarrow \frac{1}{2} (1 - \cos \theta) \cos 2 \phi \cos \omega u_o ,
  \\
  r_\mathrm{lum}^{(1)}/r \rightarrow (1 - \cos \theta) \cos 2 \phi (\cos \omega u_s - \frac{1}{2} \cos \omega u_o),
\end{eqnarray}
when $\langle h_{ij} \rangle$ can be ignored.

Continuing these calculations through second order, it is implied by \eqref{gamma2Cont} that the dominant contribution to the waveform is
	\begin{equation}
  		\bg_{(2)} (u) = -\frac{1}{8} (\omega u)^2 \bm{I} + \ldots
  		\label{gamma2ContMax}
	\end{equation}
	at large $u$. Substituting this into \eqref{Redshift2Large} results in
	\begin{equation}
  		\omega_o^{(2)}/\omega_s = \frac{ \pi^2 N^2 }{ 4 } \left[ 2 \left( \frac{ u_o }{ u_o-u_s } \right) - 1 \right] (1 - \cos \theta)  + \ldots .
  			\label{Redshift2Mono}
	\end{equation}
The $\pi^2 N^2$ factor appearing here can be enormous at large distances, potentially allowing the magnitude of $\epsilon \omega_o^{(2)}$ to compete with $\omega_o^{(1)}$. The temporal and angular dependencies of the first and second-order effects are very different, however. Similar comments also apply to the other observables considered here.

Consider, for example, $\theta_{(2)}$. This involves the second-order average $\langle \bg_{(2)} \rangle$, which is not generically negligible at any distance. Computing it using \eqref{gamma2ContMax} and substituting the result into \eqref{theta2Large} shows that
	\begin{eqnarray}
  		\theta_{(2)} &=& \frac{ \pi^2 N^2}{12 }  \bigg\{ \left[ 3 \left( \frac{u_o}{ u_o-u_s } \right) - 1 \right] (1 + \cos \theta) 
  	\nonumber	
	\\
	&& \qquad ~ - 3 \left( \frac{ u_o }{ u_o - u_s} \right)^2  \cos \theta \bigg\} \sin \theta + \ldots ,
\end{eqnarray}
which is again proportional to $\pi^2 N^2$. Although the first-order observables oscillate for monochromatic waves, their second-order counterparts act (at least over short observation times) more like offsets. These offsets depend in a characteristic way on both the distance to a source and its angular separation from the wave propagation direction. Their magnitudes can be comparable to first-order effects when $N \sim \omega r \sim \epsilon^{-1/2}$.

\subsection{Optics and the memory effect}
\label{Sect:Burst2}

As a second example, consider a short burst of gravitational waves as described in Section \ref{Sect:Burst1}. We do not model the burst itself, but only its memory in the form of the second-order Rosen waveform \eqref{gammaApproxMem}. Observations may then be split into three main phases. These are the i) early times where $u_o < -\delta$, ii) intermediate times where $u_o > 0$ but $u_s < -\delta$, and iii) late times where $u_o>0$ and $u_s > 0$. The first and last of these phases involve light propagating entirely through flat regions of spacetime.

We start by considering first-order effects. The frequency shift is particularly simple, vanishing at both early and late times, while holding the constant value
\begin{equation}
  \omega_o^{(1)}/\omega_s = - \frac{1}{2} (1-\cos \theta) \cos 2 \phi 
  \label{Omega1Mem}
\end{equation}
at intermediate times. The true frequency shift would not, of course, jump between these possibilities instantaneously. Transitions would instead last for observation times of order the burst time $\delta$, and would depend on detailed properties of the waveform. Also note that \eqref{Omega1Mem} suggests that $\omega_o^{(1)}$ does not vanish as $\theta_{(0)} \rightarrow \pi$. This is in conflict with the general comments following \eqref{Redshift1Alt}, and is an unphysical artifact of the discontinuity introduced in the waveform by ignoring timescales of order $\delta$.

Optical observables other than the frequency shift depend on $\langle h_{ij} \rangle$, which cannot necessarily be ignored when memory effects are significant. Indeed, this average imparts various observables with the ``continuous component'' of their time dependence. For example, \eqref{us1} implies that $u_s^{(0)}$ is initially zero, changes linearly with $u_o$ via
\begin{equation}
  u_s^{(1)} = - \frac{r}{2 \sqrt{2}} \left( \frac{u_o}{ u_o - u_s } \right) \sin^2 \theta \cos 2 \phi 
  \label{us1Burst}
\end{equation}
at intermediate times, and then saturates to a constant value at late times.

Astrometric effects are somewhat more complicated. Through first order, a source which is initially stationary at $(\theta, \phi) = ( \theta_{(0)}, \phi_{(0)})$ rapidly moves to a position determined by
\begin{equation}
  \theta_{(1)} = - \frac{1}{2}  \sin \theta \cos 2 \phi, \qquad \phi_{(1)} = \frac{1}{2} \sin 2 \phi
\end{equation}
when $u_o = 0$. These perturbations then vary linearly with $u_o$ until saturating at 
\begin{equation}
 	\theta_{(1)} = \frac{1}{4}  \sin 2 \theta \cos 2 \phi, \qquad \phi_{(1)} = - \frac{1}{2} \sin 2 \phi,
\end{equation}
where they remain at late times. That the asymptotic angular perturbations are nonzero is a consequence of the finite displacement memory associated with the gravitational wave. 

The distance observables $r_\mathrm{area}$ and $r_\mathrm{lum}$ are both equal to $r$ at early times. They then suffer ``immediate'' equal and opposite first-order perturbations
\begin{eqnarray}
  r_\mathrm{lum}^{(1)} = - r_\mathrm{area}^{(1)} =  \frac{1}{2} r ( 1 - \cos \theta) \cos 2 \phi 
\end{eqnarray}
when $u_o = 0$. These perturbations subsequently vary linearly with $u_o$ until reaching
\begin{eqnarray}
  r_\mathrm{area}^{(1)} = \frac{1}{2} r \sin^2 \theta \cos 2 \phi, \qquad 
  r_\mathrm{lum}^{(1)} = r_\mathrm{area}^{(1)} + r ( 1 - \cos \theta) \cos 2 \phi 
\end{eqnarray}
when $u_o$ approaches $r (1 + \cos \theta_{(0)})/\sqrt{2}$ from below. The perturbation to the area distance retains this value for all later times, while the perturbation to the luminosity distance instead jumps so as to agree (again) with $r^{(1)}_\mathrm{area}$ at late times.

Continuing these calculations through second order, we focus on the contribution due to the constant $\tr (\bm{b}_{\mathcal{F}_+})_{(2)}$ appearing in \eqref{gammaApproxMem}. As explained following \eqref{b2}, this constant must be negative if the first-order memory is to perturb only the asymptotic positions of test particles, and not their velocities. The relevant portion $\bg_{(2)}(u) = u [\tr (\bm{b}_{\mathcal{F}_+})_{(2)}] \bm{I} + \ldots$ of the second-order metric then contributes 
\begin{equation}
  \omega_o^{(2)}/\omega_s = - \frac{1}{2} u_o [\tr (\bm{b}_{\mathcal{F}_+})_{(2)}] ( 1 - \cos \theta) + \ldots 
\end{equation}
to the frequency shift at intermediate times. This is nowhere negative, and therefore corresponds to a linearly-increasing blueshift. It is an effect which saturates at
\begin{equation}
  \omega_o^{(2)}/\omega_s = - \frac{r}{2 \sqrt{2}} [\tr (\bm{b}_{\mathcal{F}_+})_{(2)}] \sin^2 \theta + \ldots,
\end{equation}
where it remains for all late times. Gravitational wave bursts therefore induce  permanent blueshifts at this order: Initially-comoving test particles are focused by the wave, experiencing a small ``velocity memory'' at late times. This is a nonlinear effect proportional to the nominal source-observer distance $r$.

Similarly considering the dominant large-distance effect on $\theta$ at second order, it follows from \eqref{theta2Large} that $\theta_{(2)}$ continuously changes from zero according to
\begin{equation}
\theta_{(2)} = \frac{1}{4} u_o [\tr (\bm{b}_{\mathcal{F}_+})_{(2)}] \left[ \left( \frac{ u_o }{ u_o - u_s } \right) ( 1 + \cos \theta) - 2 \right] \sin \theta + \ldots
\end{equation}
at intermediate times. This is nowhere negative, indicating that second-order effects tend to make objects appear to bunch up against the wave propagation direction. This behavior transitions to
\begin{equation}
  \fl \qquad \theta_{(2)} = \frac{1}{4} (u_o-u_s) [\tr (\bm{b}_{\mathcal{F}_+})_{(2)}] \left[ 2 \left( \frac{ u_o }{ u_o-u_s } \right) \cos \theta - (1 + \cos \theta) \right] \sin \theta + \ldots 
\end{equation}
at late times, the sign of which depends on $\theta_{(0)}$ and $u_o$. The nontrivial late-time dependence of $\theta$ on $u_o$ is another consequence of the velocity memory imparted by the gravitational wave burst at second order.

\subsection{Superimposed bursts and continuous waves}
\label{Sect:BurstAndCont}

Our last example consists of a superposition of the monochromatic and burst-type waves discussed above. Allowing the relative amplitudes of these components to differ by a constant factor $h_\infty$, suppose that except in a small neighborhood of $u=0$,
\begin{equation}
  \bm{h} ( u ) = [ h_\infty \Theta(u) - \cos \omega u ] 
        \left( \begin{array}{cc}
            1    &    0\\
            0    &    -1
        \end{array} \right),
    \label{hMono2}
\end{equation}
where $\Theta(u)$ is the unit step function. All first-order perturbations to the optical observables in this case have the form $(\mbox{results of Section \ref{Sect:ExMono})}+ h_\infty \times (\mbox{results of Section \ref{Sect:Burst2}})$.

Second-order effects are potentially more interesting. In particular, the $\dot{h}_{ij}$ contributions to the frequency shift \eqref{Redshift2Large} and the luminosity distance \eqref{rLum2Large} can be significant. Using \eqref{us1Burst} while assuming that the number of cycles \eqref{NDef} associated with the continuous component of the wave satisfies $N \gg 1$, this reduces to
\begin{equation}
	\frac{ u_s^{(1)} }{ r}  \left( \frac{ r \dot{h}_{ij}(u_s) \hat{k}^i_\perp \hat{k}^j_\perp }{ 1 + \cos \theta } \right) = - \pi N h_\infty ( 1 - \cos \theta)^2 \cos^2 2 \phi \sin \omega u_s
\end{equation}
at late times. Unlike the other second-order examples considered here, this oscillates with the same frequency as the first-order waveform. The prefactor $\pi N h_\infty$ can also be extremely large, particularly if the memory amplitude is much larger than the amplitude of the monochromatic component so $h_\infty \gg 1$.

\section{Conclusions}

We have derived the exact time delays \eqref{TimeDelay}, frequency shifts \eqref{RedshiftGeneral}, sky positions \eqref{kHat}, area distances \eqref{dAng1}, and luminosity distances \eqref{dLum} associated with optical observations in the presence of arbitrarily-varying gravitational plane waves. Together, these results provide a simple and non-perturbative framework with which to explore the physics of nonlinear gravitational waves.

One conclusion is that the optical effects associated with gravitational plane waves  appear particularly simple when those waves are described in terms of $\xi_{ij}(u)$, a nonlocal variable which generalizes the familiar waveforms of TT-gauge perturbation theory. Non-perturbatively, $\xi_{ij}$ is a square root of the transverse metric $\gamma_{ij} = \xi_{ki} \xi_{kj}$ in a Rosen-type coordinate system. It also represents the nontrivial components of a Jacobi propagator, and therefore describes separations between neighboring families of geodesics. The physical character of $\xi_{ij}$ can be understood by noting that it satisfies an ordinary differential equation $\ddot{\xi}_{ij} = H_{ik} \xi_{kj}$ which also describes a set of coupled parametric oscillators. The instantaneous natural frequencies of these oscillators directly correspond to those curvature components $H_{ij}(u) = -\xi^{-1}_{ki} \xi^{-1}_{lj} R_{ukul}$ which represent the freely-specifiable degrees of freedom associated with the gravitational wave. Although the equation satisfied by $\xi_{ij}$ is linear, its solutions depend nonlinearly on the ``local waveform'' $H_{ij}$. It is through this nonlinearity---which is more closely connected to the geodesic equation than to Einstein's equation---that interesting optical effects can arise on large scales.

Much of our discussion examines these effects perturbatively. An expansion $\xi_{ij} = \xi^{(0)}_{ij} + \epsilon \xi_{ij}^{(1)} + \epsilon^2 \xi_{ij}^{(2)} + \ldots$ for the metric square root is obtained in Section \ref{Sect:perts}, where $\epsilon$ is a small parameter which controls only the overall scale of $H_{ij}$. The various optical effects considered non-perturbatively in Section \ref{Sect:Optics} are then specialized in Section \ref{Sect:ObservablePerts} for freely-falling sources and observers, and also expanded through second order in $\epsilon$. Two main results emerge: i) Higher-order perturbations secularly grow at large source-observer distances, and ii) some higher-order metric perturbations produce observable effects with angular dependencies which are completely different from those associated with first-order effects.

The first of these statements can be understood directly from \eqref{H2ndOrderAlt}, which shows that the second-order metric perturbation involves the first-order metric perturbation $h_{ij}$ via a double integral of the ``energy density'' $\dot{h}_{ij} \dot{h}_{ij}$. This density is non-negative, so its integrals---and therefore the metric itself---grow wherever $\dot{h}_{ij} \neq 0$. As a consequence, second-order optical effects   can be much more important on large scales than naive estimates might suggest.

This can be interpreted as a kind of memory effect when the curvature is significant only for a short time (i.e., for gravitational wave bursts). Standard assumptions have long been known to imply that at first order, bursts can exhibit a displacement-type memory, but no ``velocity memory;'' initially-comoving particles remain comoving after a gravitational wave has passed. We show that this picture changes at second order in the gravitational wave amplitude. The secularly-growing portion of the second-order metric physically corresponds to a finite kick imparted to initially-comoving test particles, a nontrivial velocity memory. Even a small effect of this sort can produce significant displacements over sufficiently long times. If a particular burst is characterized as $N$ oscillations with angular frequency $\omega$ and strain amplitude $\epsilon$, the associated nonlinear effects on observables are fractionally of order $\epsilon^2 N (\omega r)$ over lengthscales of order $r \gg N \omega^{-1}$. 

The analogous scaling is different for continuous gravitational waves which maintain their amplitudes over all relevant lengthscales. Second-order effects are then shown to grow like $\epsilon^2 N^2$, where $N \sim \omega r$ now denotes the number of gravitational wave cycles between a source and its observer. This number is typically of order unity or less for gravitational waves intended to be observed using standard interferometer designs such as LIGO, so nonlinear effects are negligible in those cases. Much larger values of $N$ can arise in pulsar timing, however. Consider, for example, a gravitational wave with $\omega/2 \pi = 300 \, \mathrm{nHz}$ and with an approximately constant amplitude between the earth and a pulsar where $r \sim 10 \, \mathrm{kpc}$. Second-order terms in this case are then amplified by $N^2 \sim 10^{12}$. Although this factor is large, multiplying it by a realistic strain magnitude results in an $O(\epsilon^2 N^2)$ effect which would still be challenging to detect.

Regardless of the precise type of gravitational wave under consideration, we show that the growing higher-order metric perturbations also have nonvanishing traces. While the familiar trace-free property of TT gauge can be extended to all odd-order metric perturbations, even orders generically acquire finite traces. This distinction physically results in different azimuthal dependencies for the optical effects associated with, e.g., first and second-order metric perturbations. Focusing only on the aforementioned second-order terms which grow at large distances, many second-order observables in general relativity appear similar to linear observables, but with $h_{ij}$ replaced by an ``effective metric perturbation'' which possesses a small nonzero trace. This trace has optical effects similar to those associated with breathing-type polarization modes in other theories of gravity. It differs, however, in that nonlinear perturbations in general relativity do not represent physically independent degrees of freedom; such terms are entirely determined by the ordinary $+$ and $\times$ polarization modes. Furthermore, this effect arises only for nonlocal measurements. Both distinctions can be observationally subtle, however, and might complicate efforts \cite{NonEinstein} to constrain alternative theories of gravity via the presence of additional gravitational wave polarization modes.

%While the genuine first-order $h_{ij}$ has vanishing trace in vacuum general relativity, this is not necessarily true in other theories of gravity. For those theories where first-order traces can be identified with a physically-independent degree of freedom, that degree of freedom is referred as a ``breathing''-type polarization mode \cite{AltGrav}.

\ack

I thank Stanislav Babak for posing the questions which eventually led to this work, and also for a number of useful discussions along the way.

\end{document}